\documentclass{aastex631}

\pdfoutput=1
\usepackage{graphicx}
\usepackage{pgffor}
\usepackage{float}

\begin{document}

\shorttitle{Massive SF in WR Galaxies}
\shortauthors{Ferraro et al.}

\graphicspath{{./}{figures/}}

\title{VLA 22 GHz Imaging of Massive Star Formation in Local Wolf-Rayet Galaxies}

\author[0000-0002-9064-4592]{Nicholas G. Ferraro}
\affiliation{UCLA Department of Physics and Astronomy, Los Angeles, CA 90095, USA}

\author[0000-0003-4625-2951]{Jean L. Turner}
\affiliation{UCLA Department of Physics and Astronomy, Los Angeles, CA 90095, USA}

\author[0000-0002-5770-8494]{Sara C. Beck}
\affiliation{Tel Aviv University, Tel Aviv, Israel}

\author[0000-0003-3926-1139]{Edwin Alexani}
\affiliation{Rochester Institute of Technology, Rochester, NY 14623, USA}

\author[0009-0007-3025-8693]{Runa Indrei}
\affiliation{UCLA Department of Physics and Astronomy, Los Angeles, CA 90095, USA}

\author[0009-0004-1435-566X]{Bethany M. Welch}
\affiliation{UCLA Department of Physics and Astronomy, Los Angeles, CA 90095, USA}

\author[0009-0008-8444-6988]{Tunhui Xie}
\affiliation{UCLA Department of Physics and Astronomy, Los Angeles, CA 90095, USA}

\begin{abstract}
We present 22 GHz imaging of regions of massive star formation within the Local Wolf-Rayet Galaxy Sample (LWRGS), a NSF's Karl G. Jansky Very Large Array (VLA) survey of 30 local galaxies showing spectral features of Wolf-Rayet (WR) stars. These spectral features are present in galaxies with young super star clusters (SSCs), and are an indicator of large concentrations of massive stars. We present a catalog of 92 individually-identified regions of likely free-free emission associated with potential young SSCs located in these WR galaxies. The free-free fluxes from these maps allow extinction-free estimates of the Lyman continuum rates, masses, and luminosities of the emission regions. 39 of these regions meet the minimum Lyman continuum rate to contain at least once SSC, and 29 of these regions could contain individual SSCs massive enough to test specific theories on star formation and feedback inhibition in SSCs, requiring follow-up observations at higher spatial resolution. The resulting catalog provides sources for future molecular line and infrared studies into the properties of super star cluster formation.

\end{abstract}


\section{Introduction} \label{sec:intro}

Super star clusters (SSCs) are the largest observed star clusters (typically defined by $M \gtrsim 10^{5.5} M_{\odot}$), contain thousands of O stars tightly packed within diameters on order of $\sim$ 5 pc, and are likely precursors to globular clusters (GCs) \citep{2008A&A...489..567B}. Clusters with masses this high may experience a uniquely dynamic star-forming environment as a result of the extreme physical conditions created in these regions. \cite{2004ApJ...610..226S} has shown that clusters with masses $M \gtrsim 10^{6} M_{\odot}$ experience rapid mass loss of super-solar metallicity gas, rapidly self-enriching their cluster environments. These highly enriched environments, with corresponding higher mass densities, cause stellar winds and even shocks from supernovae \citep{2015ApJ...814L...8T} to immediately become radiative, quickly losing energy in the process and thus stalling within the cluster. This process, known as ``catastrophic cooling'', is caused by the injection of super-solar metallicity material from a large concentration of O stars formed on Myr timescales. This in turn inhibits the feedback of the cluster, allowing star formation to continue and the cluster to grow to even larger masses and metallicities in a self-feeding cycle.

To test these theories, we require sufficiently massive clusters in nearby galaxies, larger than any currently found in the Milky Way.
An obstacle in testing these theories of SSC-specific star formation remains the relatively small sample size of the rare clusters that are massive enough to trigger ``catastrophic cooling'', close enough to isolate with current telescope limitations, and young enough to still retain their natal gas. 
There are UV, optical and H$\alpha$ surveys of young clusters 
\citep[e.g.][]{2015AJ....149...51C,2022MNRAS.512.1294H,2023MNRAS.524.1191T}, which
provide large samples of young clusters in the local universe, but these clusters 
may suffer extinction, which can be difficult to diagnose in the UV and optical alone 
\citep[e.g.,][]{1979ApJ...228..112M, 1984ApJ...287..228N}.

Thermal free-free (Bremsstrahlung) radio continuum emission provides one of the most accurate tracers of the high mass star formation expected in SSCs. Radio continuum emission gives an extinction-free measure of the Lyman continuum rate, which allows for robust estimation of massive star formation luminosities, under the assumption of ionization boundedness. In conjunction with hydrogen recombination lines, it can be used to estimate extinction. Thermal free-free emission is also not indiscriminate towards lower mass star formation like mid-IR flux, thus can distinguish massive clusters with many O stars from collections of lower mass clusters.

To optimize our chances of detecting young massive clusters, we employ
as a beacon, another tracer of recent high mass star formation, Wolf-Rayet (WR) stars. WR stars have progenitor masses $M > 25 M_{\odot}$ and up to $\sim$ 110 $M_{\odot}$, but mean evolved masses $M \sim 9 M_{\odot}$ \citep{doi:10.1146/annurev.astro.45.051806.110615}. WR stars represent a high mass loss phase of the most massive stellar evolution with stellar ages of 3 $\sim$ 5 Myr \citep{2008A&A...485..657B}. 
Galaxies with sufficient numbers of WR stars take on their characteristic spectral features \citep[namely high excitation lines of He II $\lambda$4686, NIII $\lambda$4640, and CIII/CIV $\lambda$4650;][]{1991ApJ...377..115C}. Most extremely massive stars, the WR progenitors, are located in correspondingly massive clusters, thus WR galaxies are very likely to host clusters massive enough to be SSCs. This is not a perfect indicator of the presence of massive clusters; clusters that are too young to have massive stars or that are completely embedded with no leakage of WR signatures will be missed. However, we posit that these galaxies are likely to have such clusters.

We therefore present the Local Wolf-Rayet Galaxy Survey (LWRGS) with the goal of detecting a robust sample of young local SSCs of known mass. We use thermal free-free emission from the associated HII regions to produce an atlas of 22 GHz sources and a catalog of positions, fluxes, Lyman continuum rates, and preliminary cluster mass and luminosity estimates for SSC nebula candidates. Since thermal radio continuum emission is relatively weak, any notable detections of significant S/N at 22 GHz are very likely to be SSC nebulae. 

The paper is laid out in the following manner: In \S 2, we describe the sample selection, VLA observations, data reduction, and interferometric imaging processes for our 22 GHz radio data. \S 2 also covers the ancillary 22 $\mu$m mid-IR data included in this paper, our region identification process, and the aperture photometry techniques used on both the 22 GHz and 22 $\mu$m data. In \S 3 we present the results of the previously described processes and calculate the Lyman continuum rates and other derived properties for our identified regions. In \S 4, we discuss the population of identified SSC candidates and mid-IR/radio flux ratios. \S 5 contains the conclusions for this paper and the Appendix contains the remainder of the 22 GHz atlas.

\section{Sample and Data Analysis} \label{sec:data}

\subsection{Sample Selection} \label{sec:sample}
\begin{deluxetable}{lccc}[!ht]

\tablecaption{LWRGS Galaxy Properties}
\tablenum{1}

\tablehead{\colhead{Galaxy} & \colhead{Type$^{a}$} & \colhead{Dist.$^{b}$} & \colhead{D$_{25}^{a}$} \\ 
\colhead{} & \colhead{} & \colhead{(Mpc)} & \colhead{(arcmin)} } 

\startdata
NGC 1156 & IB(s)m & 7.6 & 3.3 x 2.5 \\
UGC 02866 & Sc & 16.7 & 1.0 x 0.9 \\
Mrk 86 & SB(s)m pec & 13.7 & 1.7 x 1.5 \\
NGC 2541 & SA(s)cd & 12.6 & 6.3 x 3.2 \\
NGC 3003 & Sbc* & 19.5 & 5.8 x 1.3 \\
NGC 3265 & E* & 25.3 & 1.3 x 1.0 \\
Mrk 33 & Im pec* & 15.4 & 1.0 x 0.9 \\
NGC 3310 & SAB(r)bc pec & 18.7 & 3.1 x 2.4 \\
NGC 3353 & Sb pec* & 20.1 & 1.3 x 1.0 \\
NGC 3423 & SA(s)cd & 11.7 & 3.8 x 3.2 \\
NGC 3451 & Sd & 28.1 & 1.7 x 0.8 \\
IC 0691 & I? & 23.7 & 0.6 x 0.4 \\
Mrk 1450 & I & 16.3 & 0.5 x 0.4 \\
Mrk 750 & cI & 13.6 & 0.5 x 0.3 \\
Mrk 1307 & Pec & 20.7 & 0.6 x 0.5 \\
Mrk 1308 & S0 & 18.9 & 0.7 x 0.6 \\
NGC 4216 & SAB(s)b* & 14.5 & 8.1 x 1.8 \\
NGC 4236 & SB(s)dm & 4.4 & 21.9 x 7.2 \\
NGC 4369 & (R)SA(rs)a & 21.6 & 2.1 x 2.0 \\
NGC 4389 & SB(rs)bc pec* & 7.5 & 2.6 x 1.3 \\
NGC 4449 & IBm & 4.0 & 6.2 x 4.4 \\
NGC 4490 & SB(s)d pec & 6.5 & 6.3 x 3.1 \\
NGC 4656 & SB(s)m pec & 7.9 & 15.1 x 3.0 \\
NGC 4670 & SB(s)0/a pec & 19.6 & 1.4 x 1.1 \\
NGC 4691 & (R)SB(s)0/a pec & 20.6 & 2.8 x 2.3 \\
NGC 4713 & SAB(rs)d & 13.7 & 2.7 x 1.7 \\
NGC 4808 & SA(s)cd* & 17.5 & 2.8 x 1.1 \\
NGC 4861 & SB(s)m* & 10.0 & 4.0 x 1.5 \\
NGC 4900 & SB(rs)c & 26.4 & 2.2 x 2.1 \\
M 83 & SAB(s)c & 4.7 & 12.9 x 11.5 \\
\enddata
\tablenotetext{a}{Morphological types and diameters taken from the Third Reference Catalog of Bright Galaxies (RC3:\cite{1991rc3..book.....D})}
\tablenotetext{b}{Distances taken from NASA/IPAC Extragalactic Database}
\label{tab:gal_prop}
\end{deluxetable}
The complete LWRGS sample comprises new observations of 35 pointings taken in 30 local (d $\lesssim$ 30 Mpc) WR galaxies. All galaxies are initially chosen from one of two WR galaxy catalogs, \cite{1999A&AS..136...35S} and \cite{2008A&A...485..657B}, containing 139 and 570 WR galaxies respectively. Examining the confirmed SSC within NGC 5253, NGC 5253-D1 \citep[][]{,2017ApJ...846...73T}, provides the foundation for several selection criteria. Accounting for the necessity of molecular line followup observations with additional telescopes (\emph{ALMA}, \emph{NOEMA}, \emph{SMA}), we apply a distance limit of $\sim$ 20 Mpc, set by the approximately maximal distance at which ALMA would detect CO(3-2) with 2 km/s resolution.
We apply minimum IR fluxes (taken from WISE and IRAS) at 12 and 24 $\mu$m (2 and 8.8 Jy) to remove sources without sufficient IR emission to host a single SSC. We also remove galaxies known to host AGN as to avoid ambiguity of corresponding emission features with WR features, leaving $\sim$80 galaxies. All galaxies for which VLA K (18 - 26.5 GHz) or Ka (26.5 - 40 GHz) band data taken in C or D configuration exists within the NRAO archive are removed for this sample. These archival galaxies will be added to the sample, expanding the catalog and atlas in future work. We also remove any galaxies too southerly for observation with the VLA. The final 30 selected galaxies are listed in Table 1, along with their morphologies, distances, and diameters.

\subsection{VLA Observations and Data Reduction} \label{sec:reduction}
Observations in the VLA K band (18 - 26.5 GHz) were taken during both the 2021 and 2022 VLA C-configuration cycles using the 3-bit, 2 GHz samplers for maximum bandwidth and better continuum sensitivity. The observations (project codes 21A-246 and 22B-071) utilized the full 8 GHz bandwidth of the receivers, with baseband centers located at 19, 21, 23, and 25 GHz to minimize overlapping and maximize coverage. We used the IR flux of NGC 5253 at 24 $\mu m$ (8.8 Jy) to estimate the expected median 22 GHz flux density for the source list, and utilized this number to set the minimum time on source for each target as $\sim$4 minutes. C-Configuration was chosen for this survey because of the FWHM (full-width half maximum) of the synthesized beam/angular resolution of 0.95” for K band observations. 

At a distance of 20 Mpc, 1\arcsec\ corresponds to a size of 100 pc, on the scale of a giant molecular cloud (GMC). Thus while this survey’s resolution is too low to resolve a 5 pc SSC, it is high enough to isolate and identify large HII regions, possibly containing multiple SSCs contributing to the observed star formation. 

Scheduling blocks used the standard VLA flux density calibrators 3C 147 and 3C 286. Table 2 contains observation dates and times, pointing center coordinates, and complex gain/phase calibrators selected for each target. Antenna reference pointing calibration was performed every $\sim$30 minutes in accordance with VLA recommendations for the band/configuration combination. Complex gain calibration scans occurred before and after each target source, within a recommended six minute cumulative cycle time.

VLA data reduction was performed using the Common Astronomy Software Applications \citep[CASA;][]{2007ASPC..376..127M} version 6.1.2, utilizing the VLA continuum optimized calibration pipeline. Additional adjustments to the pipeline were made by NRAO data analysts as part of the Science Ready Data Products (SRDP) initiative, ensuring image ready data products after final inspections of visibilities and calibration tables for any remaining radio frequency interference (RFI) or hardware artifacts.

\begin{deluxetable*}{lccccc}[!ht]

\tablecaption{VLA Observations}
\tablenum{2}

\tablehead{\colhead{Galaxy} & \colhead{Date} & \colhead{Time} & \colhead{RA (J2000)} & \colhead{Dec (J2000)} & \colhead{Phase Calibrator} \\ 
\colhead{(Pointing)} & \colhead{} & \colhead{(UTC)} & \colhead{(hh:mm:ss)} & \colhead{(dd:mm:ss)} & \colhead{} } 

\startdata
Mrk 86 & 2021Jun22 & 17:59:39.0 & 08:13:13.05 & +45:59:26.5 & J0920+4441 \\
NGC 2541 &  & 18:14:36.0 & 08:14:37.28 & +49:03:00.1 & J0920+4441 \\
NGC 3003 &  & 18:30:33.0 & 09:48:36.505 & +33:25:20.68 & J0958+3224 \\
UGC 02866 & 2021Jun23 & 12:26:24.0 & 03:50:14.890 & +70:05:40.90 & J0410+7656 \\
NGC 1156 &  & 12:38:24.0 & 02:59:42.546 & +25:14:15.01 & J0237+2848 \\
NGC 3353 & 2021Sep04 & 16:33:27.0 & 10:45:21.978 & +55:57:38.87 & J0937+5008 \\
Mrk 33 &  & 16:39:27.0 & 10:32:31.964 & +54:24:03.69 & J0937+5008 \\
NGC 3310 &  & 16:45:24.0 & 10:38:44.85 & +53:30:05.2 & J0937+5008 \\
NGC 3265 &  & 16:59:24.0 & 10:31:06.77 & +28:47:48.0 & J1014+2301 \\
IC 0691 & 2021Sep06 & 18:34:15.0 & 11:26:44.30 & +59:09:19.6 & J1153+4931 \\
Mrk 1450 &  & 18:40:12.0 & 11:38:35.661 & +57:52:26.98 & J1153+4931 \\
Mrk 750 &  & 18:52:57.0 & 11:50:02.642 & +15:01:23.06 & J1224+2122 \\
NGC 4216 &  & 18:58:54.0 & 12:15:54.250 & +13:08:59.89 & J1224+2122 \\
Mrk 1307 &  & 19:09:24.0 & 11:52:37.452 & -02:28:09.53 & J1222+0413 \\
Mrk 1308 &  & 19:15:24.0 & 11:54:12.27 & +00:08:11.7 & J1222+0413 \\
NGC 4369 & 2021Sep14 & 15:48:06.0 & 12:24:35.78 & +39:22:53.7 & J1146+3958 \\
NGC 4389 &  & 15:54:03.0 & 12:25:34.60 & +45:41:06.8 & J1146+3958 \\
NGC 4449 &  & 16:00:03.0 & 12:28:12.016 & +44:05:40.94 & J1146+3958 \\
NGC 4449 N &  & 16:06:03.0 & 12:28:15.100 & +44:07:00.00 & J1146+3958 \\
NGC 4236 S &  & 16:17:30.0 & 12:16:47.000 & +69:27:05.00 & J1400+6210 \\
NGC 4236 C &  & 16:23:30.0 & 12:16:38.400 & +69:29:03.00 & J1400+6210 \\
NGC 4236 N &  & 16:29:30.0 & 12:16:15.900 & +69:30:37.00 & J1400+6210 \\
NGC 4490 E & 2022Oct01 & 16:27:15.0 & 12:30:37.43 & +41:38:28.0 & J1146+3958 \\
NGC 4490 W &  & 16:33:15.0 & 12:30:31.90 & +41:39:10.0 & J1146+3958 \\
NGC 4656 &  & 15:58:18.0 & 12:43:56.69 & +32:10:14.7 & J1310+3220 \\
NGC 4670 &  & 16:04:18.0 & 12:45:16.982 & +27:07:30.39 & J1310+3220 \\
NGC 4861 S &  & 16:10:18.0 & 12:59:00.800 & +34:51:07.50 & J1310+3220 \\
NGC 4861 N &  & 16:16:18.0 & 12:59:03.800 & +34:52:30.00 & J1310+3220 \\
NGC 4691 & 2022Nov07 & 16:12:39.0 & 12:48:13.63 & -03:19:57.8 & J1239+0730 \\
NGC 4713 &  & 16:18:36.0 & 12:49:58.15 & +05:18:54.7 & J1239+0730 \\
NGC 4808 &  & 16:24:36.0 & 12:55:48.40 & +04:18:20.0 & J1239+0730 \\
NGC 4900 &  & 16:30:36.0 & 13:00:39.25 & +02:30:02.7 & J1239+0730 \\
M 83 &  & 16:42:33.0 & 13:37:00.461 & -29:51:55.00 & J1316-3338 \\
NGC 3451 & 2022Sep30 & 15:20:06.0 & 10:54:21.87 & +27:14:22.2 & J1014+2301 \\
NGC 3423 C &  & 15:37:00.0 & 10:51:14.45 & +05:50:21.9 & J1041+0610 \\
\enddata

\label{tab:vla_obs}

\end{deluxetable*}

\subsection{Interferometric Imaging} \label{sec:imaging}
Each source was imaged using the CASA version 6.1.2 task TCLEAN applied to the final calibrated VLA measurement sets. TCLEAN’s spectral definition mode was set to multi-frequency synthesis \citep[MFS;][]{1990MNRAS.246..490C,1994A&AS..108..585S} for continuum imaging and the Multi-term (Multi Scale) Multi-Frequency Synthesis \citep{2011A&A...532A..71R} deconvolver algorithm was used to account for different angular scales of extended sources, set to 0, 1, 2, and 3 times the FWHM of the synthesized beam, and up to half the minor axis scale of the mapped structure if necessary for the select few extended sources, within the largest angular scale of 66\arcsec. The pixel size was set to 0.19\arcsec\ so there would be 5 pixels across the 0.95\arcsec\ synthesized beam and the image size was set to 600x600 pixels to account for the $\sim$ 2\arcmin\ diffraction-limited field of view (EVLA Memo 195). We used Briggs weighting (with ROBUST values ranging from 0.5 to 1.0) to balance imaging sensitivity with resolution.

Additional imaging parameters, such as stopping threshold and small scale bias, were optimized for the best compromise between brightness, artifact reduction, and maximal S/N. The CASA task WIDEBANDPBCOR was applied to each TCLEAN-produced image to correct for the primary beam and improve the accuracy of calculated flux densities from the images.

\subsection{Ancillary Data} \label{sec:ancillary}
Mid-IR continuum fluxes for the sample were taken from NASA's Wide-field Infrared Survey Explorer \citep[WISE:][]{2010AJ....140.1868W} AllWISE Source Catalog. WISE data in the W4 band (centered at 22 $\mu$m) have an angular resolution of $\sim$ 12.0\arcsec, roughly 140 times larger in area than the VLA's synthesized beam, and have 5$\sigma$ point source sensitivity of 6 mJy. Analytical parameters for WISE Source Catalog photometry are taken from section 2.3 of the Explanatory Supplement to the WISE All-Sky Data Release Products \citep{2012wise.rept....1C}.

\begin{figure*}[!htp]
    \centering
    \includegraphics[width=100mm,scale=0.5]{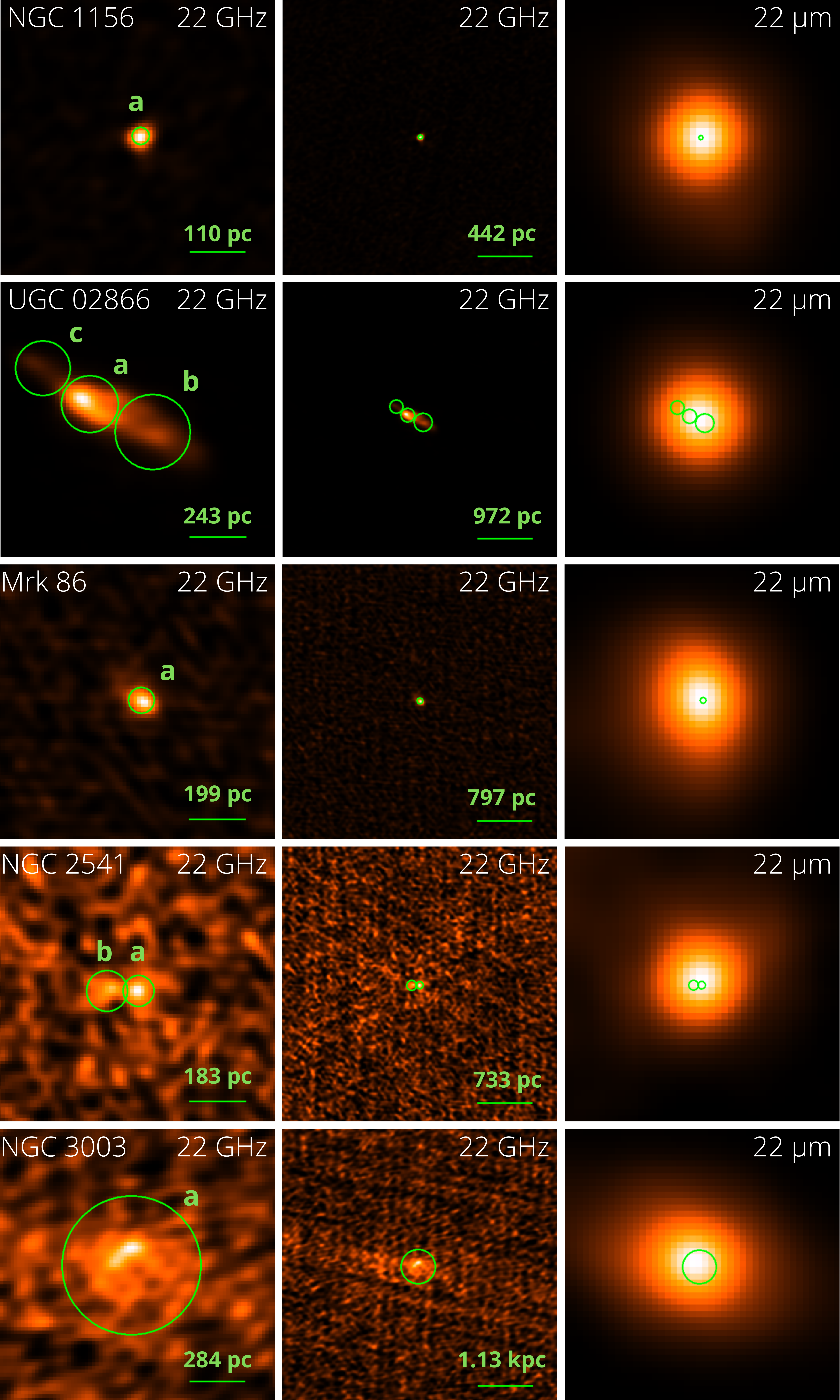}
    \caption{3-panel images for 5 of the 30 galaxies in the LWRGS, where each row shows two maps of our VLA 22 GHz observations, the central 15\arcsec\ map (left) and the larger 1.0\arcmin\ map (center), with its corresponding WISE 22 $\mu$m 1.0' map (right). We used a squared stretch color scale and include scale bars in each 22 GHz map, of lengths 3.0\arcsec\ and 12\arcsec, for the 15\arcsec\ and 1.0\arcmin\ maps respectively. The labeled regions correspond with the regions in Tables 2, 3, and 4, while region sizes reflect the 2D Gaussian fitting-determined image component size. Figures 3, 4, 5, 6, 7, and 8, in the Appendix show the remainder of the 92 individually identified star-forming regions in our survey.}
    \label{fig:a1}
\end{figure*}

\subsection{Region Identification and Aperture Photometry} \label{sec:regid}
The relatively small nature of our VLA images at 22 GHz ($\sim$ 2\arcmin) allowed for the direct visual identification of regions with strong sources of radio K band flux, i.e. regions of potential high-mass star formation. Using the CASA image viewer application, some initially identified regions of 22 GHz flux were split into multiple individual regions based on peaks and variations in flux density across the apparent region. Within the 35 successfully reduced pointings, 92 individual regions were identified. All pointings contained at least one visually-identified region, with the exception of Mrk 1308, NGC 4236 S, NGC 4389, NGC 4861 N, NGC 4713, and NGC 4900 that all lacked detections of significant S/N ($\lesssim$ 3). All individual regions were given an internal identifier (`Atlas ID') containing the name of their host galaxy plus a lower case letter suffix. In sources with multiple regions of 22 GHz flux, the region with the highest peak flux density was designated with the suffix `a', the region with the next highest peak flux density was assigned `b', and so forth. Regions identified in galaxies containing only one region were simply given the suffix `a'. Regions were also given a more specific external identifier (`Catalog ID') based on coordinates of the peak flux density.

Using the Image Fitting Dialog within the Cube Analysis and Rendering Tool for Astronomy \citep[CARTA;][]{angus_comrie_2018_3377984}, we fit two-dimensional Gaussians to each individual region to measure the peak flux density (in mJy beam$^{-1}$) and integrated flux (in mJy). We used the CASA task IMFIT to measure image component size deconvolved with the synthesized beam (in arcsec) if calculable. We measured rms around the flux region using the CASA task IMSTAT within the CASA image viewer application. The results are in Table 3.

\section{Results} \label{sec:results}

\subsection{Region Flux Measurements}
Figure 1, along with Figures 3, 4, 5, 6, 7, and 8 in the Appendix, show both VLA 22 GHz maps and the WISE 22 $\mu$m maps for 30 of our pointings that altogether contain 92 individual regions of significant 22 GHz flux. Radio continuum is typically fairly weak so any identified regions are likely to be the result of aggregate emission from star-forming regions containing clusters of various size, including SSCs. Some figures contain multiple sets of maps to show all identified regions within the pointing. In all VLA 15.0" maps, each identified region is labeled using a lowercase letter to match its Atlas ID and encircled in an aperture reflecting the approximate image component size (major axis) given by the IMFIT output shown in Table 3. These apertures are projected onto the WISE 22 $\mu$m maps, demonstrating that the regions of significant 22 GHz flux are generally coincident with the 22 $\mu$m flux. This is expected because while radio continuum flux is a good tracer of specifically high-mass star formation, mid-IR flux is a tracer of general star formation, including high-mass star formation.

The large beam size of WISE also means it is more likely to pick up flux coming from nearby regions, including extended regions of low mass star formation that will not have 22 GHz free-free flux. This is most notable in comparing the VLA and WISE 1.0' maps in the pointings of Figures 1, 3, 4, 5, 6, 7, and 8 with multiple 22 GHz-identified regions, such as UGC 02866 in row 2 of Figure 1. Here, the importance of radio continuum, and specifically radio continuum observations with sub-arcsecond resolution, in detecting smaller, individual star-forming regions is clearly shown.

Table 3 gives the Atlas ID, peak flux coordinates (RA and declination in J2000), peak flux density, rms, integrated flux, and image component size (deconvolved with the synthesized beam), with the addition of the size, and position angle, of the synthesized beam used in the corresponding TCLEAN run for the pointing. Photometric uncertainties are determined by the CARTA Image Fitting tool.

\startlongtable
\begin{deluxetable*}{lcccccccc}

\tablecaption{Region Photometry \& Parameters}
\tablenum{3}

\tablehead{\colhead{Atlas ID} & \colhead{Peak RA$^{a}$} & \colhead{Peak dec$^{a}$} & \colhead{Peak Flux} & \colhead{rms$^{b}$} & \colhead{Synthesized} & \colhead{PA$^{c}$} & \colhead{Integrated} & \colhead{Component} \\ 
\colhead{} & \colhead{} & \colhead{} & \colhead{Density} & \colhead{} & \colhead{Beam} & \colhead{} & \colhead{Flux} & \colhead{Size$^{d}$} \\
\colhead{} & \colhead{(hh:mm:ss)} & \colhead{(dd:mm:ss)} & \colhead{(mJy bm$^{-1}$)} & \colhead{} & \colhead{} & \colhead{($^{\circ}$)} & \colhead{(mJy)} & \colhead{(deconvolved)} } 

\startdata
NGC 1156 a & 02:59:41.28 & 25:14:15.05 & 1.04 $\pm $ 0.11 & 0.026 & 1.24'' x 1.05'' & -61.1 & 1.18 $\pm $ 0.21 & 0.5'' x 0.4'' \\
        UGC 02866 a & 03:50:15.39 & 70:05:41.64 & 2.57 $\pm $ 0.54 & 0.1 & 1.55'' x 0.91'' & 74.2 & 16.44 $\pm $ 3.98 & 3.0'' x 1.6'' \\
        UGC 02866 b & 03:50:14.75 & 70:05:39.88 & 1.30 $\pm $ 0.26 & 0.1 & 1.55'' x 0.91'' & 74.2 & 11.09 $\pm $ 2.49 & 4.0'' x 1.7'' \\
        UGC 02866 c & 03:50:15.96 & 70:05:43.77 & 0.90 $\pm $ 0.32 & 0.025 & 1.55'' x 0.91'' & 74.2 & 2.06 $\pm $ 1.00 & 2.9'' x 1.0'' \\
        Mrk 86 a & 08:13:12.98 & 45:59:38.58 & 0.21 $\pm $ 0.03 & 0.014 & 1.23'' x 0.97'' & 62.1 & 0.41 $\pm $ 0.10 & 1.4'' x 0.9'' \\
        NGC 2541 a & 08:14:37.29 & 49:02:59.69 & 0.11 $\pm $ 0.02 & 0.02 & 1.25'' x 0.98'' & 53.4 & 0.17 $\pm $ 0.05 & 1.7'' x 1.3'' \\
        NGC 2541 b & 08:14:37.46 & 49:02:59.73 & 0.08 $\pm $ 0.02 & 0.02 & 1.25'' x 0.98'' & 53.4 & 0.19 $\pm $ 0.08 & 2.2'' x 0.4'' \\
        NGC 3003 a & 09:48:35.70 & 33:25:17.47 & 0.14 $\pm $ 0.02 & 0.021 & 1.56'' x 1.08'' & -80.6 & 1.10 $\pm $ 0.17 & 7.5'' x 4.5'' \\
        NGC 3265 a & 10:31:06.78 & 28:47:48.04 & 0.17 $\pm $ 0.04 & 0.016 & 1.23'' x 1.00'' & -1.7 & 0.23 $\pm $ 0.08 & 2.2'' x 1.2'' \\
        Mrk 33 a & 10:32:31.97 & 54:24:02.34 & 0.57 $\pm $ 0.14 & 0.050 & 1.29'' x 0.95'' & 28.5 & 1.57 $\pm $ 0.52 & 2.3'' x 1.1'' \\
        Mrk 33 b & 10:32:31.81 & 54:24:04.09 & 0.40 $\pm $ 0.12 & 0.050 & 1.29'' x 0.95'' & 28.5 & 1.02 $\pm $ 0.41 & 2.5'' x 1.0'' \\
        NGC 3310 a & 10:38:45.86 & 53:30:11.87 & 0.98 $\pm $ 0.12 & 0.030 & 1.28'' x 0.95'' & 28.2 & 1.16 $\pm $ 0.23 & 0.5'' x 0.4'' \\
        NGC 3310 b & 10:38:44.80 & 53:30:04.90 & 0.57 $\pm $ 0.17 & 0.044 & 1.28'' x 0.95'' & 28.2 & 1.81 $\pm $ 0.69 & 1.9'' x 1.2'' \\
        NGC 3310 c & 10:38:46.92 & 53:30:16.50 & 0.55 $\pm $ 0.09 & 0.026 & 1.28'' x 0.95'' & 28.2 & 1.12 $\pm $ 0.25 & 1.6'' x 0.8'' \\
        NGC 3310 d & 10:38:46.64 & 53:30:11.57 & 0.49 $\pm $ 0.12 & 0.057 & 1.28'' x 0.95'' & 28.2 & 1.45 $\pm $ 0.44 & 1.8'' x 1.1'' \\
        NGC 3310 e & 10:38:44.75 & 53:30:03.00 & 0.27 $\pm $ 0.11 & 0.044 & 1.28'' x 0.95'' & 28.2 & 0.43 $\pm $ 0.25 & 1.4'' x 0.8'' \\
        NGC 3353 a & 10:45:22.00 & 55:57:39.75 & 0.94 $\pm $ 0.23 & 0.055 & 1.34'' x 0.98'' & 35.5 & 2.95 $\pm $ 0.94 & 2.2'' x 1.3'' \\
        NGC 3353 b & 10:45:21.84 & 55:57:41.56 & 0.29 $\pm $ 0.09 & 0.055 & 1.34'' x 0.98'' & 35.5 & 0.45 $\pm $ 0.20 & 1.6'' x 1.0'' \\
        NGC 3423 a & 10:51:14.13 & 05:50:33.16 & 0.08 $\pm $ 0.02 & 0.017 & 0.93'' x 0.93'' & -25.2 & 0.12 $\pm $ 0.05 & e \\
        NGC 3423 b & 10:51:15.39 & 05:50:12.54 & 0.08 $\pm $ 0.03 & 0.022 & 1.25'' x 0.93'' & -25.2 & 0.09 $\pm $ 0.05 & 0.8'' x 0.7'' \\
        NGC 3451 a & 10:54:22.19 & 27:14:26.89 & 0.08 $\pm $ 0.02 & 0.014 & 1.05'' x 1.03'' & -52.3 & 0.12 $\pm $ 0.03 & e \\
        NGC 3451 b & 10:54:21.87 & 27:14:22.25 & 0.07 $\pm $ 0.02 & 0.016 & 1.05'' x 1.03'' & -52.3 & 0.19 $\pm $ 0.05 & 1.5'' x 1.0'' \\
        IC 0691 a & 11:26:44.21 & 59:09:21.34 & 0.28 $\pm $ 0.07 & 0.031 & 1.50'' x 0.85'' & 11.7 & 0.96 $\pm $ 0.29 & 3.1'' x 1.0'' \\
        IC 0691 b & 11:26:43.29 & 59:09:37.57 & 0.13 $\pm $ 0.04 & 0.034 & 1.50'' x 0.85'' & 11.7 & 0.50 $\pm $ 0.18 & 2.9'' x 0.5'' \\
        Mrk 1450 a & 11:38:35.65 & 57:52:27.13 & 0.12 $\pm $ 0.03 & 0.023 & 1.45'' x 0.80'' & 13.1 & 0.30 $\pm $ 0.11 & 1.5'' x 0.5'' \\
        Mrk 750 a & 11:50:02.70 & 15:01:23.21 & 0.14 $\pm $ 0.02 & 0.033 & 1.42'' x 0.83'' & -1.5 & 2.23 $\pm $ 0.31 & 4.6'' x 2.7'' \\
        Mrk 1307 a & 11:52:37.20 & -02:28:09.99 & 0.11 $\pm $ 0.03 & 0.021 & 1.64'' x 0.79'' & -6.8 & 0.12 $\pm $ 0.05 & e \\
        NGC 4216 a & 12:15:54.39 & 13:08:58.44 & 0.43 $\pm $ 0.11 & 0.028 & 1.39'' x 0.82'' & -1.5 & 0.70 $\pm $ 0.26 & e \\
        NGC 4216 b & 12:15:54.48 & 13:09:00.74 & 0.40 $\pm $ 0.12 & 0.028 & 1.39'' x 0.82'' & -1.5 & 0.54 $\pm $ 0.25 & e \\
        NGC 4216 c & 12:15:54.31 & 13:08:57.41 & 0.37 $\pm $ 0.09 & 0.028 & 1.39'' x 0.82'' & -1.5 & 0.39 $\pm $ 0.16 & e \\
        NGC 4216 d & 12:15:54.32 & 13:08:55.23 & 0.16 $\pm $ 0.05 & 0.040 & 1.39'' x 0.82'' & -1.5 & 0.18 $\pm $ 0.10 & e \\
        NGC 4216 e & 12:15:54.68 & 13:08:56.56 & 0.15 $\pm $ 0.04 & 0.032 & 1.39'' x 0.82'' & -1.5 & 0.15 $\pm $ 0.08 & e \\
        NGC 4236 a & 12:16:35.55 & 69:28:00.70 & 3.45 $\pm $ 0.88 & 0.12 & 1.19'' x 0.71'' & 78.0 & 5.32 $\pm $ 2.08 & 0.6'' x 0.4'' \\
        NGC 4236 b & 12:16:37.89 & 69:28:47.84 & 0.46 $\pm $ 0.13 & 0.025 & 1.19'' x 0.71'' & 78.0 & 0.55 $\pm $ 0.25 & 0.4'' x 0.2'' \\
        NGC 4236 c & 12:16:17.11 & 69:30:41.57 & 0.27 $\pm $ 0.08 & 0.022 & 1.12'' x 0.70'' & 72.5 & 0.40 $\pm $ 0.19 & 0.9'' x 0.3'' \\
        NGC 4236 d & 12:16:39.92 & 69:29:02.66 & 0.16 $\pm $ 0.05 & 0.026 & 1.19'' x 0.71'' & 78.0 & 0.24 $\pm $ 0.11 & 1.0'' x 0.4'' \\
        NGC 4369 a & 12:24:35.62 & 39:23:02.95 & 0.57 $\pm $ 0.15 & 0.031 & 1.11'' x 0.77'' & 89.6 & 0.80 $\pm $ 0.32 & 0.8'' x 0.6'' \\
        NGC 4449 a & 12:28:10.95 & 44:06:48.43 & 0.54 $\pm $ 0.14 & 0.079 & 1.09'' x 0.75'' & 87.8 & 0.86 $\pm $ 0.33 & e \\
        NGC 4449 b & 12:28:13.85 & 44:07:10.38 & 0.27 $\pm $ 0.04 & 0.041 & 1.09'' x 0.75'' & 87.8 & 1.74 $\pm $ 0.32 & 4.0'' x 2.7'' \\
        NGC 4449 c & 12:28:13.09 & 44:05:42.58 & 0.16 $\pm $ 0.05 & 0.028 & 0.95'' x 0.42'' & 12.6 & 0.49 $\pm $ 0.20 & 0.8'' x 0.5'' \\
        NGC 4449 d & 12:28:11.86 & 44:05:39.07 & 0.13 $\pm $ 0.05 & 0.025 & 0.95'' x 0.42'' & 12.6 & 0.26 $\pm $ 0.15 & 0.8'' x 0.1'' \\
        NGC 4449 e & 12:28:11.50 & 44:05:38.16 & 0.10 $\pm $ 0.03 & 0.027 & 0.95'' x 0.42'' & 12.6 & 0.37 $\pm $ 0.15 & e \\
        NGC 4490 a & 12:30:34.49 & 41:38:26.03 & 0.93 $\pm $ 0.25 & 0.023 & 1.16'' x 1.03'' & 58.8 & 2.62 $\pm $ 0.93 & 1.8'' x 1.0'' \\
        NGC 4490 b & 12:30:29.56 & 41:39:26.75 & 0.43 $\pm $ 0.08 & 0.056 & 1.20'' x 1.03'' & 60.6 & 1.52 $\pm $ 0.37 & 2.3'' x 1.8'' \\
        NGC 4490 c & 12:30:29.48 & 41:39:28.11 & 0.41 $\pm $ 0.09 & 0.034 & 1.20'' x 1.03'' & 60.6 & 1.42 $\pm $ 0.39 & 2.1'' x 1.6'' \\
        NGC 4490 d & 12:30:28.15 & 41:39:39.32 & 0.37 $\pm $ 0.11 & 0.038 & 1.20'' x 1.03'' & 60.6 & 0.66 $\pm $ 0.28 & 1.1'' x 0.8'' \\
        NGC 4490 e & 12:30:29.31 & 41:39:28.08 & 0.36 $\pm $ 0.06 & 0.039 & 1.20'' x 1.03'' & 60.6 & 3.04 $\pm $ 0.59 & 3.6'' x 2.9'' \\
        NGC 4490 f & 12:30:29.15 & 41:39:28.16 & 0.36 $\pm $ 0.06 & 0.039 & 1.20'' x 1.03'' & 60.6 & 2.22 $\pm $ 0.44 & 3.3'' x 2.3'' \\
        NGC 4490 g & 12:30:34.53 & 41:38:33.23 & 0.23 $\pm $ 0.06 & 0.023 & 1.16'' x 1.03'' & 58.8 & 0.26 $\pm $ 0.11 & e \\
        NGC 4490 h & 12:30:37.73 & 41:37:58.55 & 0.20 $\pm $ 0.05 & 0.014 & 1.16'' x 1.03'' & 58.8 & 0.26 $\pm $ 0.11 & e \\
        NGC 4490 i & 12:30:34.90 & 41:39:02.34 & 0.16 $\pm $ 0.05 & 0.026 & 1.16'' x 1.03'' & 58.8 & 0.22 $\pm $ 0.10 & e \\
        NGC 4490 j & 12:30:34.04 & 41:38:49.25 & 0.16 $\pm $ 0.03 & 0.024 & 1.16'' x 1.03'' & 58.8 & 0.18 $\pm $ 0.06 & e \\
        NGC 4490 k & 12:30:34.35 & 41:38:46.45 & 0.13 $\pm $ 0.04 & 0.024 & 1.16'' x 1.03'' & 58.8 & 0.13 $\pm $ 0.06 & e \\
        NGC 4490 l & 12:30:34.18 & 41:38:50.93 & 0.12 $\pm $ 0.04 & 0.021 & 1.16'' x 1.03'' & 58.8 & 0.23 $\pm $ 0.11 & e \\
        NGC 4490 m & 12:30:33.60 & 41:38:43.96 & 0.10 $\pm $ 0.03 & 0.024 & 1.16'' x 1.03'' & 58.8 & 0.11 $\pm $ 0.05 & e \\
        NGC 4490 n & 12:30:35.22 & 41:38:58.98 & 0.10 $\pm $ 0.03 & 0.024 & 1.16'' x 1.03'' & 58.8 & 0.19 $\pm $ 0.08 & e \\
        NGC 4490 o & 12:30:35.07 & 41:38:56.96 & 0.10 $\pm $ 0.03 & 0.024 & 1.16'' x 1.03'' & 58.8 & 0.15 $\pm $ 0.06 & e \\
        NGC 4490 p & 12:30:34.46 & 41:38:43.54 & 0.09 $\pm $ 0.03 & 0.028 & 1.16'' x 1.03'' & 58.8 & 0.11 $\pm $ 0.05 & e \\
        NGC 4490 q & 12:30:34.45 & 41:38:51.59 & 0.09 $\pm $ 0.03 & 0.026 & 1.16'' x 1.03'' & 58.8 & 0.21 $\pm $ 0.10 & 1.2'' x 0.5'' \\
        NGC 4490 r & 12:30:34.13 & 41:38:54.90 & 0.08 $\pm $ 0.02 & 0.024 & 1.16'' x 1.03'' & 58.8 & 0.13 $\pm $ 0.05 & 1.6'' x 0.8'' \\
        NGC 4656 a & 12:43:56.65 & 32:10:12.41 & 0.28 $\pm $ 0.07 & 0.028 & 1.30'' x 1.08'' & -83.0 & 1.06 $\pm $ 0.31 & 2.3'' x 1.8'' \\
        NGC 4656 b & 12:43:57.61 & 32:10:09.18 & 0.25 $\pm $ 0.06 & 0.018 & 1.30'' x 1.08'' & -83.0 & 0.25 $\pm $ 0.10 & e \\
        NGC 4656 c & 12:43:58.03 & 32:10:07.67 & 0.08 $\pm $ 0.01 & 0.016 & 1.30'' x 1.08'' & -83.0 & 0.08 $\pm $ 0.03 & e \\
        NGC 4670 a & 12:45:16.98 & 27:07:31.64 & 0.11 $\pm $ 0.01 & 0.018 & 1.37'' x 1.09'' & -81.3 & 0.95 $\pm $ 0.14 & 3.8'' x 1.0'' \\
        NGC 4670 b & 12:45:17.08 & 27:07:28.86 & 0.10 $\pm $ 0.03 & 0.018 & 1.37'' x 1.09'' & -81.3 & 0.13 $\pm $ 0.06 & 2.0'' x 0.9'' \\
        NGC 4670 c & 12:45:17.17 & 27:07:28.80 & 0.10 $\pm $ 0.02 & 0.018 & 1.37'' x 1.09'' & -81.3 & 0.10 $\pm $ 0.04 & 1.9'' x 1.3'' \\
        NGC 4670 d & 12:45:16.76 & 27:07:30.94 & 0.08 $\pm $ 0.01 & 0.018 & 1.37'' x 1.09'' & -81.3 & 0.25 $\pm $ 0.05 & 2.2'' x 1.3'' \\
        NGC 4691 a & 12:48:13.91 & -03:20:00.44 & 1.00 $\pm $ 0.08 & 0.022 & 1.33'' x 0.92'' & -1.1 & 1.78 $\pm $ 0.21 & 1.0'' x 0.7'' \\
        NGC 4691 b & 12:48:14.17 & -03:20:00.22 & 0.66 $\pm $ 0.14 & 0.022 & 1.33'' x 0.92'' & -1.1 & 0.77 $\pm $ 0.28 & 0.6'' x 0.3'' \\
        NGC 4691 c & 12:48:13.05 & -03:19:59.85 & 0.24 $\pm $ 0.05 & 0.012 & 1.33'' x 0.92'' & -1.1 & 0.29 $\pm $ 0.10 & e \\
        NGC 4691 d & 12:48:13.61 & -03:19:57.51 & 0.17 $\pm $ 0.04 & 0.032 & 1.33'' x 0.92'' & -1.1 & 0.23 $\pm $ 0.09 & 0.6'' x 0.5'' \\
        NGC 4691 e & 12:48:12.54 & -03:20:00.26 & 0.13 $\pm $ 0.03 & 0.014 & 1.33'' x 0.92'' & -1.1 & 0.23 $\pm $ 0.08 & 1.0'' x 0.7'' \\
        NGC 4691 f & 12:48:12.66 & -03:19:57.72 & 0.11 $\pm $ 0.03 & 0.016 & 1.33'' x 0.92'' & -1.1 & 0.37 $\pm $ 0.12 & 1.8'' x 1.5'' \\
        NGC 4691 g & 12:48:12.72 & -03:19:56.54 & 0.10 $\pm $ 0.03 & 0.015 & 1.33'' x 0.92'' & -1.1 & 0.32 $\pm $ 0.10 & 1.8'' x 0.7'' \\
        NGC 4808 a & 12:55:47.64 & 04:18:32.95 & 0.11 $\pm $ 0.02 & 0.018 & 1.17'' x 0.88'' & -8.2 & 0.52 $\pm $ 0.11 & 2.1'' x 1.6'' \\
        NGC 4808 b & 12:55:50.51 & 04:17:57.09 & 0.09 $\pm $ 0.03 & 0.023 & 1.17'' x 0.88'' & -8.2 & 0.14 $\pm $ 0.07 & 0.8'' x 0.3'' \\
        NGC 4808 c & 12:55:49.54 & 04:18:06.69 & 0.08 $\pm $ 0.02 & 0.017 & 1.17'' x 0.88'' & -8.2 & 0.19 $\pm $ 0.07 & 1.9'' x 0.7'' \\
        NGC 4861 a & 12:59:00.41 & 34:50:43.93 & 0.52 $\pm $ 0.12 & 0.016 & 1.36'' x 1.04'' & -88.5 & 1.18 $\pm $ 0.39 & 1.7'' x 1.4'' \\
        NGC 4861 b & 12:59:00.32 & 34:50:42.31 & 0.32 $\pm $ 0.06 & 0.021 & 1.36'' x 1.04'' & -88.5 & 1.49 $\pm $ 0.35 & 1.9'' x 0.9'' \\
        M 83 a & 13:37:00.34 & -29:51:50.64 & 2.10 $\pm $ 0.33 & 0.080 & 2.54'' x 0.93'' & -13.7 & 11.49 $\pm $ 2.14 & 3.8'' x 2.6'' \\
        M 83 b & 13:37:00.53 & -29:51:54.82 & 2.02 $\pm $ 0.36 & 0.080 & 2.54'' x 0.93'' & -13.7 & 11.43 $\pm $ 2.37 & 4.3'' x 1.7'' \\
        M 83 c & 13:37:00.27 & -29:51:52.79 & 1.82 $\pm $ 0.29 & 0.080 & 2.54'' x 0.93'' & -13.7 & 8.89 $\pm $ 1.67 & 3.9'' x 2.0'' \\
        M 83 d & 13:37:00.38 & -29:51:59.65 & 1.07 $\pm $ 0.19 & 0.080 & 2.54'' x 0.93'' & -13.7 & 3.18 $\pm $ 0.75 & 3.3'' x 1.5'' \\
        M 83 e & 13:36:58.37 & -29:51:04.63 & 0.84 $\pm $ 0.21 & 0.068 & 2.54'' x 0.93'' & -13.7 & 0.83 $\pm $ 0.41 & e \\
        M 83 f & 13:37:00.80 & -29:51:55.55 & 0.59 $\pm $ 0.07 & 0.080 & 2.54'' x 0.93'' & -13.7 & 3.58 $\pm $ 0.52 & 3.1'' x 2.5'' \\
        M 83 g & 13:37:01.11 & -29:51:52.30 & 0.38 $\pm $ 0.07 & 0.090 & 2.54'' x 0.93'' & -13.7 & 0.86 $\pm $ 0.24 & 2.0'' x 1.1'' \\
        M 83 h & 13:37:00.89 & -29:52:03.62 & 0.37 $\pm $ 0.05 & 0.091 & 2.54'' x 0.93'' & -13.7 & 2.09 $\pm $ 0.34 & 4.7'' x 2.1'' \\
        M 83 i & 13:37:01.06 & -29:51:58.17 & 0.36 $\pm $ 0.02 & 0.090 & 2.54'' x 0.93'' & -13.7 & 3.12 $\pm $ 0.18 & 6.1'' x 3.7'' \\
        M 83 j & 13:37:00.11 & -29:51:55.38 & 0.30 $\pm $ 0.06 & 0.080 & 2.54'' x 0.93'' & -13.7 & 2.16 $\pm $ 0.45 & 3.5'' x 2.4'' \\
        M 83 k & 13:37:00.03 & -29:52:16.38 & 0.32 $\pm $ 0.07 & 0.051 & 2.54'' x 0.93'' & -13.7 & 0.63 $\pm $ 0.21 & 1.9'' x 0.7'' \\
        M 83 l & 13:37:01.35 & -29:51:27.01 & 0.17 $\pm $ 0.04 & 0.027 & 2.54'' x 0.93'' & -13.7 & 0.25 $\pm $ 0.08 & 1.7'' x 0.4'' \\
        M 83 m & 13:36:59.81 & -29:52:25.93 & 0.14 $\pm $ 0.03 & 0.036 & 2.54'' x 0.93'' & -13.7 & 0.24 $\pm $ 0.07 & 1.0'' x 0.1'' \\
\enddata
\tablenotetext{a}{All coordinates in J2000. Peak position error is a function of beam and source structure, precision is indicated by number of significant digits reported.}
\tablenotetext{b}{Units also mJy bm$^{-1}$}
\tablenotetext{c}{Position angle}
\tablenotetext{d}{Upper limits as determined by CASA's IMFIT task}
\tablenotetext{e}{CASA unable to estimate image component size due to point-source nature}

\end{deluxetable*}
\tabletypesize{\small}

\subsubsection{Comparing Flux Measurements}
To directly compare flux densities of regions in our VLA 22 GHz maps and the coincident sources of 22 $\mu$m flux from the AllWISE Source Catalog, we need to both calculate flux measurements for the AllWISE sources based on the catalog magnitudes and also artificially match the 12.0'' resolution of WISE on the VLA maps. We calculate fluxes for the AllWISE sources using Equation 1, taken from the Explanatory Supplement to the WISE All-Sky Data Release Products \citep{2012wise.rept....1C}

\begin{equation}
    F_{\nu}[Jy] \approx 0.90 \times (8.284/f_{c}) \times 10^{(-m/2.5)}
\end{equation}
where $m$ is the corresponding magnitude of the source and $f_{c}$ is a flux correction term based on the colors of the source, ranging from 0.9907 to 1.0130.

To calculate corresponding 22 GHz fluxes, we use two different methods: summation and convolution. 
For the summation calculation, we add the integrated fluxes of the 22 GHz regions, as seen in Table 3, that would fit within the corresponding WISE source. For the convolution calculation, we use the CASA task IMSMOOTH to perform a Fourier-based convolution of the original maps with a Gaussian kernel sized to produce new 22 GHz maps with a resolution matching that of the WISE W4 band. The convolved flux can then be calculated using the IMFIT task. 
Ordinarily, convolution would be preferred; however for aperture synthesis observations, negative sidelobes affect flux measurements beyond the largest angular scale (here 66\arcsec). This ``bowl effect'' can cause differences between the convolved and summed fluxes.
The results from the WISE flux calculations, summed VLA flux calculations, convolved VLA flux calculations, and mid-IR/radio flux ratios can be found in Section 4.1 in Table 4.

\subsection{Lyman Continuum Rates} \label{sec:lcr}
Approximating all measured 22 GHz flux as thermal free-free emission \citep[accurate within an order of magnitude;][]{2018ApJS..234...24M} allows us to measure the Lyman continuum rates of each identified region using Equation 2 presented in \cite{2016ApJ...833L...6C},
\begin{equation}
    N_{Lyc} = 1.25 \times 10^{50} T_{4}^{-0.51} \nu_{11}^{0.118} (\frac{n_{i}}{n_{p}}) D_{Mpc}^{2} S(mJy)
\end{equation}
assuming that $T = 10000 K$ and $\frac{n_{i}}{n_{p}} = 1$. We then use the following relations derived from a 2016 STARBURST 99 simulation \citep{1999ApJS..123....3L}, assuming continuous rotating star formation of $0.3 M_{\odot}/yr$, a full Kroupa IMF (with $0.1 M_{\odot} \lesssim M \lesssim 150 M_{\odot}$), and metallicty $Z = 0.008$,
\begin{equation}
    log(M/M_{\odot}/N_{Lyc}) = -46.67
\end{equation}
\begin{equation}
    log(L/L_{\odot}/N_{Lyc}) = -43.70
\end{equation}
to calculate the estimated mass (Equation 3) and luminosity (Equation 4) of the possible SSC nebulae within each region. Table 4 contains the measured Lyman continuum rate, estimated mass, and estimated luminosity for each identified region. The median Lyman continuum rate, mass and luminosity of the identified regions are $1.76 \times 10^{52} s^{-1}$, $3.76 \times 10^{5} M_{\odot}$ and $3.51 \times 10^{8} L_{\odot}$, respectively. 

While we typically define SSCs as having $N_{Lyc} \gtrsim 10^{52} s^{-1}$, the largest Galactic clusters, found in the W49N starburst region, have $N_{Lyc} \sim 10^{50} s^{-1}$ \citep[][]{2000ApJ...540..308D,2003ApJ...589L..45A}. Thus while Table 4 is divided based on the Lyman continuum rate, all identified regions have sufficient measured $N_{Lyc}$ so that they can be classified at least as possible host regions for massive clusters (including SSCs) with varying degrees of confidence. It is important to note that while mass estimates are fairly model dependent, luminosity estimates based on Lyman continuum rates are robust. Combining this benefit with the extinction-free nature of 22 GHz flux produces an excellent method of measuring aggregate cluster luminosities.

\startlongtable
\begin{deluxetable*}{lcccccc}
\tablecaption{Star Forming Region Properties of LWRGS Atlas Regions}
\tablenum{4}

\tablehead{\colhead{Catalog ID} & \colhead{Atlas ID} & \colhead{$S_{22 GHz}$} & \colhead{$N_{Lyc}$} & \colhead{Mass} & \colhead{Luminosity} & \colhead{Maximum}\\ 
\colhead{} & \colhead{} & \colhead{} & \colhead{} & \colhead{} & \colhead{} & \colhead{Size Estimate$^{a}$}\\
\colhead{(`VLAK '...)} & \colhead{} & \colhead{(mJy)} & \colhead{($10^{52} s^{-1}$)} & \colhead{($10^{6} M_{\odot}$)} & \colhead{($10^{8} L_{\odot}$)} & \colhead{(pc)}  } 

\startdata
\hline \\
\multicolumn{6}{c}{Star Forming Regions with $N_{Lyc} \gtrsim 10^{52} s^{-1}$} \\
\hline \\
 J035015.39+700541.6 & UGC 02866 a & 16.44 $\pm $ 3.98 & 47.98 & 10.26 & 95.73 & 276 \\
        J035014.75+700539.9 & UGC 02866 b & 11.09 $\pm $ 2.49 & 32.38 & 6.92 & 64.60 & 351 \\
        J035015.96+700543.8 & UGC 02866 c & 2.06 $\pm $ 1.00 & 6.02 & 1.29 & 12.02 & 251 \\
        J094835.70+332517.5 & NGC 3003 a & 1.10 $\pm $ 0.17 & 4.37 & 0.93 & 8.72 & 824 \\
        J103106.78+284748.0 & NGC 3265 a & 0.23 $\pm $ 0.08 & 1.51 & 0.32 & 3.02 & 302 \\
        J103231.97+542402.3 & Mrk 33 a & 1.57 $\pm $ 0.52 & 3.90 & 0.83 & 7.79 & 194 \\
        J103231.81+542404.1 & Mrk 33 b & 1.02 $\pm $ 0.41 & 2.54 & 0.54 & 5.07 & 199 \\
        J103845.86+533011.9 & NGC 3310 a & 1.16 $\pm $ 0.23 & 4.23 & 0.90 & 8.43 & 57 \\
        J103844.80+533004.9 & NGC 3310 b & 1.81 $\pm $ 0.69 & 6.62 & 1.42 & 13.22 & 206 \\
        J103846.92+533016.5 & NGC 3310 c & 1.12 $\pm $ 0.25 & 4.10 & 0.88 & 8.18 & 159 \\
        J103846.64+533011.6 & NGC 3310 d & 1.45 $\pm $ 0.44 & 5.29 & 1.13 & 10.55 & 194 \\
        J103844.75+533003.0 & NGC 3310 e & 0.43 $\pm $ 0.25 & 1.58 & 0.34 & 3.15 & 144 \\
        J104522.00+555739.7 & NGC 3353 a & 2.95 $\pm $ 0.94 & 12.47 & 2.67 & 24.89 & 249 \\
        J104521.84+555741.6 & NGC 3353 b & 0.45 $\pm $ 0.20 & 1.92 & 0.41 & 3.82 & 182 \\
        J105421.87+271422.3 & NGC 3451 b & 0.19 $\pm $ 0.05 & 1.59 & 0.34 & 3.17 & 246 \\
        J112644.21+590921.3 & IC 0691 a & 0.96 $\pm $ 0.29 & 5.66 & 1.21 & 11.28 & 377 \\
        J112643.29+590937.6 & IC 0691 b & 0.50 $\pm $ 0.18 & 2.91 & 0.62 & 5.81 & 341 \\
        J115002.70+150123.2 & Mrk 750 a & 2.23 $\pm $ 0.31 & 4.31 & 0.92 & 8.61 & 351 \\
        J121554.39+130858.4 & NGC 4216 a & 0.70 $\pm $ 0.26 & 1.54 & 0.33 & 3.08 & 67 \\
        J121554.48+130900.7 & NGC 4216 b & 0.54 $\pm $ 0.25 & 1.18 & 0.25 & 2.35 & 67 \\
        J121635.55+692800.7 & NGC 4236 a & 5.32 $\pm $ 2.08 & 1.08 & 0.23 & 2.15 & 15 \\
        J122435.62+392302.9 & NGC 4369 a & 0.80 $\pm $ 0.32 & 3.90 & 0.83 & 7.78 & 105 \\
        J123034.49+413826.0 & NGC 4490 a & 2.62 $\pm $ 0.93 & 1.16 & 0.25 & 2.32 & 66 \\
        J123029.31+413928.1 & NGC 4490 e & 3.04 $\pm $ 0.59 & 1.35 & 0.29 & 2.68 & 145 \\
        J124516.98+270731.6 & NGC 4670 a & 0.95 $\pm $ 0.14 & 3.81 & 0.81 & 7.61 & 373 \\
        J124516.76+270730.9 & NGC 4670 d & 0.25 $\pm $ 0.05 & 1.02 & 0.22 & 2.04 & 242 \\
        J124813.91-032000.4 & NGC 4691 a & 1.78 $\pm $ 0.21 & 7.90 & 1.69 & 15.76 & 120 \\
        J124814.17-032000.2 & NGC 4691 b & 0.77 $\pm $ 0.28 & 3.42 & 0.73 & 6.83 & 72 \\
        J124813.05-031959.9 & NGC 4691 c & 0.29 $\pm $ 0.10 & 1.29 & 0.28 & 2.58 & 95 \\
        J124813.61-031957.5 & NGC 4691 d & 0.23 $\pm $ 0.09 & 1.03 & 0.22 & 2.06 & 81 \\
        J124812.54-032000.3 & NGC 4691 e & 0.23 $\pm $ 0.08 & 1.01 & 0.22 & 2.02 & 118 \\
        J124812.66-031957.7 & NGC 4691 f & 0.37 $\pm $ 0.12 & 1.66 & 0.36 & 3.31 & 234 \\
        J124812.72-031956.5 & NGC 4691 g & 0.32 $\pm $ 0.10 & 1.40 & 0.30 & 2.80 & 196 \\
        J125547.64+041832.9 & NGC 4808 a & 0.52 $\pm $ 0.11 & 1.68 & 0.36 & 3.34 & 226 \\
        J125900.41+345043.9 & NGC 4861 a & 1.18 $\pm $ 0.39 & 1.23 & 0.26 & 2.46 & 104 \\
        J125900.32+345042.3 & NGC 4861 b & 1.49 $\pm $ 0.35 & 1.56 & 0.33 & 3.11 & 100 \\
        J133700.34-295150.6 & M 83 a & 11.49 $\pm $ 2.14 & 2.66 & 0.57 & 5.30 & 105 \\
        J133700.53-295154.8 & M 83 b & 11.43 $\pm $ 2.37 & 2.64 & 0.56 & 5.27 & 105 \\
        J133700.27-295152.8 & M 83 c & 8.89 $\pm $ 1.67 & 2.06 & 0.44 & 4.10 & 100 \\
\hline \\
\multicolumn{6}{c}{Star Forming Regions with $N_{Lyc} \lesssim 10^{52} s^{-1}$} \\
\hline \
 J025941.28+251415.1 & NGC 1156 a & 1.18 $\pm $ 0.21 & 0.71 & 0.15 & 1.42 & 23 \\
        J081312.98+455938.6 & Mrk 86 a & 0.41 $\pm $ 0.10 & 0.80 & 0.17 & 1.60 & 112 \\
        J081437.29+490259.7 & NGC 2541 a & 0.17 $\pm $ 0.05 & 0.27 & 0.06 & 0.55 & 128 \\
        J081437.46+490259.7 & NGC 2541 b & 0.19 $\pm $ 0.08 & 0.32 & 0.07 & 0.63 & 135 \\
        J105114.13+055033.2 & NGC 3423 a & 0.12 $\pm $ 0.05 & 0.17 & 0.04 & 0.33 & 54 \\
        J105115.39+055012.5 & NGC 3423 b & 0.09 $\pm $ 0.05 & 0.12 & 0.03 & 0.25 & 63 \\
        J105422.19+271426.9 & NGC 3451 a & 0.12 $\pm $ 0.03 & 0.98 & 0.21 & 1.95 & 129 \\
        J113835.65+575227.1 & Mrk 1450 a & 0.30 $\pm $ 0.11 & 0.83 & 0.18 & 1.65 & 127 \\
        J115237.20-022810.0 & Mrk 1307 a & 0.12 $\pm $ 0.05 & 0.52 & 0.11 & 1.05 & 95 \\
        J121554.31+130857.4 & NGC 4216 c & 0.39 $\pm $ 0.16 & 0.85 & 0.18 & 1.70 & 67 \\
        J121554.32+130855.2 & NGC 4216 d & 0.18 $\pm $ 0.10 & 0.39 & 0.08 & 0.78 & 67 \\
        J121554.68+130856.6 & NGC 4216 e & 0.15 $\pm $ 0.08 & 0.34 & 0.07 & 0.67 & 67 \\
        J121637.89+692847.8 & NGC 4236 b & 0.55 $\pm $ 0.25 & 0.11 & 0.02 & 0.22 & 9 \\
        J121617.11+693041.6 & NGC 4236 c & 0.40 $\pm $ 0.19 & 0.08 & 0.02 & 0.16 & 20 \\
        J121639.92+692902.7 & NGC 4236 d & 0.24 $\pm $ 0.11 & 0.05 & 0.01 & 0.10 & 23 \\
        J122810.95+440648.4 & NGC 4449 a & 0.86 $\pm $ 0.33 & 0.14 & 0.03 & 0.29 & 18 \\
        J122813.85+440710.4 & NGC 4449 b & 1.74 $\pm $ 0.32 & 0.29 & 0.06 & 0.58 & 94 \\
        J122813.09+440542.6 & NGC 4449 c & 0.49 $\pm $ 0.20 & 0.08 & 0.02 & 0.16 & 18 \\
        J122811.86+440539.1 & NGC 4449 d & 0.26 $\pm $ 0.15 & 0.04 & 0.01 & 0.09 & 15 \\
        J122811.50+440538.2 & NGC 4449 e & 0.37 $\pm $ 0.15 & 0.06 & 0.01 & 0.12 & 18 \\
        J123029.56+413926.7 & NGC 4490 b & 1.52 $\pm $ 0.37 & 0.67 & 0.14 & 1.34 & 92 \\
        J123029.48+413928.1 & NGC 4490 c & 1.42 $\pm $ 0.39 & 0.63 & 0.13 & 1.25 & 83 \\
        J123028.15+413939.3 & NGC 4490 d & 0.66 $\pm $ 0.28 & 0.29 & 0.06 & 0.58 & 43 \\
        J123029.15+413928.2 & NGC 4490 f & 2.22 $\pm $ 0.44 & 0.98 & 0.21 & 1.96 & 127 \\
        J123034.53+413833.2 & NGC 4490 g & 0.26 $\pm $ 0.11 & 0.12 & 0.02 & 0.23 & 30 \\
        J123037.73+413758.5 & NGC 4490 h & 0.26 $\pm $ 0.11 & 0.12 & 0.02 & 0.23 & 30 \\
        J123034.90+413902.3 & NGC 4490 i & 0.22 $\pm $ 0.10 & 0.10 & 0.02 & 0.19 & 30 \\
        J123034.04+413849.3 & NGC 4490 j & 0.18 $\pm $ 0.06 & 0.08 & 0.02 & 0.16 & 30 \\
        J123034.35+413846.4 & NGC 4490 k & 0.13 $\pm $ 0.06 & 0.06 & 0.01 & 0.12 & 30 \\
        J123034.18+413850.9 & NGC 4490 l & 0.23 $\pm $ 0.11 & 0.10 & 0.02 & 0.20 & 30 \\
        J123033.60+413844.0 & NGC 4490 m & 0.11 $\pm $ 0.05 & 0.05 & 0.01 & 0.09 & 30 \\
        J123035.22+413859.0 & NGC 4490 n & 0.19 $\pm $ 0.08 & 0.09 & 0.02 & 0.17 & 30 \\
        J123035.07+413857.0 & NGC 4490 o & 0.15 $\pm $ 0.06 & 0.07 & 0.01 & 0.13 & 30 \\
        J123034.46+413843.5 & NGC 4490 p & 0.11 $\pm $ 0.05 & 0.05 & 0.01 & 0.10 & 30 \\
        J123034.45+413851.6 & NGC 4490 q & 0.21 $\pm $ 0.10 & 0.09 & 0.02 & 0.18 & 41 \\
        J123034.13+413854.9 & NGC 4490 r & 0.13 $\pm $ 0.05 & 0.06 & 0.01 & 0.11 & 56 \\
        J124356.65+321012.4 & NGC 4656 a & 1.06 $\pm $ 0.31 & 0.69 & 0.15 & 1.38 & 112 \\
        J124357.61+321009.2 & NGC 4656 b & 0.25 $\pm $ 0.10 & 0.16 & 0.03 & 0.32 & 36 \\
        J124358.03+321007.7 & NGC 4656 c & 0.08 $\pm $ 0.03 & 0.05 & 0.01 & 0.10 & 36 \\
        J124517.08+270728.9 & NGC 4670 b & 0.13 $\pm $ 0.06 & 0.52 & 0.11 & 1.03 & 207 \\
        J124517.17+270728.8 & NGC 4670 c & 0.10 $\pm $ 0.04 & 0.41 & 0.09 & 0.82 & 219 \\
        J125550.51+041757.1 & NGC 4808 b & 0.14 $\pm $ 0.07 & 0.46 & 0.10 & 0.91 & 75 \\
        J125549.54+041806.7 & NGC 4808 c & 0.19 $\pm $ 0.07 & 0.60 & 0.13 & 1.19 & 170 \\
        J133700.38-295159.6 & M 83 d & 3.18 $\pm $ 0.75 & 0.74 & 0.16 & 1.47 & 82 \\
        J133658.37-295104.6 & M 83 e & 0.83 $\pm $ 0.41 & 0.19 & 0.04 & 0.38 & 22 \\
        J133700.80-295155.5 & M 83 f & 3.58 $\pm $ 0.52 & 0.83 & 0.18 & 1.65 & 91 \\
        J133701.11-295152.3 & M 83 g & 0.86 $\pm $ 0.24 & 0.20 & 0.04 & 0.40 & 52 \\
        J133700.89-295203.6 & M 83 h & 2.09 $\pm $ 0.34 & 0.48 & 0.10 & 0.96 & 117 \\
        J133701.06-295158.2 & M 83 i & 3.12 $\pm $ 0.18 & 0.72 & 0.15 & 1.44 & 162 \\
        J133700.11-295155.4 & M 83 j & 2.16 $\pm $ 0.45 & 0.50 & 0.11 & 0.99 & 97 \\
        J133700.03-295216.4 & M 83 k & 0.63 $\pm $ 0.21 & 0.14 & 0.03 & 0.29 & 46 \\
        J133701.35-295127.0 & M 83 l & 0.25 $\pm $ 0.08 & 0.06 & 0.01 & 0.11 & 41 \\
        J133659.81-295225.9 & M 83 m & 0.24 $\pm $ 0.07 & 0.06 & 0.01 & 0.11 & 22 \\
\enddata
\tablenotetext{a}{Maximum size estimate calculated using host galaxy distance from Table 1 and major and minor axes of deconvolved image size from Table 3.}

\end{deluxetable*}

\section{Discussion} \label{sec:discussion}

\subsection{SSC Candidates} \label{sec:ssc_candidates}
As shown in Table 4, all 92 identified regions have Lyman continuum rates at least an order of magnitude greater than those of the largest clusters found within the Milky Way. 39 of these identified regions have $N_{Lyc} > 10^{52} s^{-1}$, matching the stricter criterion for an HII region associated with at least one SSC. Of these 39 most luminous star forming regions within the LWRGS sample, 29 also have estimated masses ($M \gtrsim 10^{5.5} M_{\odot}$) large enough for us to designate them as possible regions in which to study the ``catastrophic'' cooling mechanisms theorized by \cite{10.1093/mnras/staa705} and consequent stalling of supernova remnants (SNR) hypothesized by \cite{2015ApJ...814L...8T}.  If even only half of these regions are confirmed to host SSCs in follow up molecular line observations, we will have significantly increased the number of known young local clusters sufficiently massive enough to study these theories.

\subsection{Possible Non-Thermal Emission} \label{sec:NTE}
The high frequency and high angular resolution of these observations eliminates most sources of nonthermal synchrotron emission \citep{1994ApJ...421..122T, 2020ApJS..248...25L}. Two galaxies in the LWRGS sample, NGC 4236 and UGC 02866, are possible examples of such nonthermal emission. 

\subsubsection{NGC 4236}
It is possible, although unlikely, to detect individual SNR in very nearby galaxies. Both CasA \citep{2015ApJ...805..119O} and the Crab nebula \citep{2009MNRAS.396..365H} would have 22 GHz fluxes of $S_{22GHz} = 4\sim5 mJy$ $D(Mpc)^{-2}$; they could be detected in the sample out to $\sim$ 8 Mpc. NGC 4236, located only 4.4 Mpc away, is host to one of the brighter star-forming regions in the LWRGS sample, NGC 4236 a (Catalog ID: J121635.54+0692800.7). Though the region has a calculated Lyman continuum rate of $\sim 10^{52}$ s$^{-1}$ as noted in Table 4, the corresponding WISE source was very dim in mid-IR, producing the lowest calculated mid-IR/radio flux ratio of all star-forming regions in Table 5. This appears to demonstrate nonthermal emission, arising not from star formation but perhaps a SNR, requiring further observations to confidently classify the region.

\subsubsection{UGC 02866}
One galaxy, UGC 02866, has extended emission at 22 GHz, which could be either extended and very luminous massive star formation, or very bright synchrotron emission. Two regions, UGC 02866 a and UGC 02866 b (Catalog IDs J035015.40+0700541.7 and 035014.75+0700540.2 respectively), are noted in Table 4 as having the highest calculated Lyman continuum rates, an order of magnitude greater than the fluxes of the next brightest sources. We also note that the 22 GHz map of UGC 02886 shows extended emission (see Figure 1, row 2), not seen in the rest of the galaxies of our sample (Figures 3, 4, 5, 6, 7, and 8). This deviation from the rest of the sample leads us to hypothesize that a large percentage of the 22 GHz flux measured for UGC 02866 is non-thermal synchrotron emission, not thermal free-free emission. This would affect all quantities derived from the integrated flux listed in Table 4, including Lyman continuum rate and cluster mass/luminosity estimates. \cite{2020ApJS..248...25L} demonstrate that $\gtrsim$ 90$\%$ of Ka flux in local galaxies is thermal (Bremmstrahlung) but future lower frequency followup observations of UGC 02866 would be necessary to measure spectral indices and determine what percentage of 22 GHz flux in UGC 02866 is in fact thermal vs. nonthermal. Also of note, while the VLA synthesized beam size is considerably smaller than that of WISE, the $\sim$ 1.0" beam still corresponds to a resolution of $\sim$ 84 pc. A larger beam is inherently more likely to pick up extended emission, like synchrotron emission, meaning followup studies with the VLA at higher resolution are also an option to better isolate and identify any HII regions specifically within UGC 02866.

\subsection{Mid-IR/Radio Flux Ratio} \label{sec:fluxratio}
Our radio and mid-IR flux calculations in Table 5 allow us to compute mid-IR/radio ratios for each corresponding AllWISE source. This is specifically a ratio of mid-IR continuum, known to be well correlated with emission in HII regions, to free-free emission, rather than the synchrotron emission often included in galaxy scale radio-IR correlations \citep{2022A&A...667A..30V,2021MNRAS.504..118M,2019MNRAS.482..560M}.
Figure 2 shows the distribution of flux ratios for each identified region, plotted (with corresponding uncertainty measurements) using the convolved 22 GHz fluxes and using the summed 22 GHz fluxes respectively. We perform a simple ordinary least squares (OLS) linear regression on the distributions to find the lines of best fit and measure approximate relations of $S_{22 GHz, mJy} \approx 3.34$ $\times S_{22 \mu m, Jy} + 2.22$ Jy and $S_{22 GHz, mJy} \approx 3.43$ $\times S_{22 \mu m, Jy} + 0.39$ Jy for the convolved and summed radio fluxes.

We find Pearson correlation coefficients for the convolved and summed fluxes of 0.896 and 0.924 respectively, demonstrating strong linear associations between the mid-IR and radio fluxes for these star-forming regions.
Our median calculated ratio for the summed radio fluxes (218.8) agrees with the ratio observed in NGC 5253 ($S_{22 \mu m}$ = 8.82 Jy, $S_{22GHz}$ = 40.0 mJy) \citep{2001ApJ...554L..29G} though our median calculated ratio for the convolved radio fluxes disagrees by a factor of $\sim$3. It is worth noting that the flux measurements for NGC 5253 are taken over the entire galaxy, comprising much larger scales and containing multiple clusters. The relatively large spread of ratios, over multiple orders of magnitude, shown in Figure 2 and Table 5 further demonstrates that higher resolution IR data are required for more accurate comparison of mid-IR and radio fluxes, proving that WISE data are not a viable way to find star forming regions containing SSCs.

We choose a flux ratio of $S_{22 \mu m}$/$S_{22 GHz}$ $\approx 220$ going forward, defaulting to the median summed flux ratio. We use this quantity in conjunction with Equation 2 to derive Equation 5,
\begin{equation}
    N_{Lyc} = 4.76 \times 10^{50} T_{4}^{-0.51} (\frac{n_{i}}{n_{p}}) D_{Mpc}^{2} S_{22 \mu m}(Jy)
\end{equation}
allowing for the calculation of Lyman continuum rates from mid-IR flux. Using Equations 3 and 4, we can also derive Equations 6 and 7, to calculate cluster mass and luminosity estimates respectively from mid-IR flux.

\begin{equation}
    \frac{M}{M_{\odot}} = 1.02 \times 10^{4} T_{4}^{-0.51}  (\frac{n_{i}}{n_{p}}) D_{Mpc}^{2} S_{22 \mu m}(Jy)
\end{equation}

\begin{equation}
   \frac{L}{L_{\odot}} = 9.49 \times 10^{6} T_{4}^{-0.51} (\frac{n_{i}}{n_{p}}) D_{Mpc}^{2} S_{22 \mu m}(Jy)
\end{equation}

\startlongtable
\begin{deluxetable*}{lcccccc}

\tablecaption{Corresponding WISE Sources and mid-IR/Radio Flux Ratios}
\tablenum{5}

\tablehead{\colhead{WISE Source} & \colhead{Survey ID} & \colhead{$S_{22 \mu m}$} & \colhead{$S_{22GHz}$} & \colhead{$S_{22GHz}$} & \colhead{$S_{22 \mu m}/S_{22 GHz}$} & \colhead{$S_{22 \mu m}/S_{22 GHz}$} \\ 
\colhead{} & \colhead{} & \colhead{} & \colhead{Convolved} & \colhead{Summed} & \colhead{Convolved} & \colhead{Summed} \\
\colhead{} & \colhead{(`VLAK '...)} & \colhead{(Jy)} & \colhead{(mJy)} & \colhead{(mJy)} & \colhead{} & \colhead{} } 

\startdata
J025941.30+251415.0 & J025941.28+251415.1 & 0.30 $\pm $ 0.01 & 1.41 $\pm $ 0.25 & 1.18 $\pm $ 0.21 & 209.3 & 250.2 \\
J035015.06+700541.0 & J035015.39+700541.6 & 4.75 $\pm $ 0.04 & 26.99 $\pm $ 0.91 & 29.59 $\pm $ 4.80 & 175.9 & 160.4 \\
~ & J035014.75+700539.9 & ~ & ~ & ~ & ~ & ~ \\
~ & J035015.96+700543.8 & ~ & ~ & ~ & ~ & ~ \\
J081313.07+455939.8 & J081312.98+455938.6 & 0.15 $\pm $ 0.004 & 0.81 $\pm $ 0.08 & 0.41 $\pm $ 0.10 & 189.6 & 372.2 \\
J081437.27+490259.7 & J081437.29+490259.7 & 0.03 $\pm $ 0.001 & 2.33 $\pm $ 0.38 & 0.36 $\pm $ 0.09 & 11.3 & 73.3 \\
~ & J081437.46+490259.7 & ~ & ~ & ~ & ~ & ~ \\
J094835.79+332517.8 & J094835.70+332517.5 & 0.10 $\pm $ 0.003 & 7.44 $\pm $ 0.38 & 1.10 $\pm $ 0.17 & 14.1 & 95.3 \\
J103106.77+284747.9 & J103106.78+284748.0 & 0.24 $\pm $ 0.01 & 0.59 $\pm $ 0.09 & 0.23 $\pm $ 0.08 & 413.7 & 1,057.8 \\
J103231.91+542403.0 & J103231.97+542402.3 & 0.72 $\pm $ 0.01 & 3.79 $\pm $ 0.56 & 2.59 $\pm $ 0.66 & 188.9 & 276.4 \\
~ & J103231.81+542404.1 & ~ & ~ & ~ & ~ & ~ \\
J103846.03+533011.8 & J103845.86+533011.9 & 4.93 $\pm $ 0.05 & 10 $\pm $ 1.1 & 5.97 $\pm $ 0.92 & 492.6 & 825.1 \\
~ & J103844.80+533004.9 & ~ & ~ & ~ & ~ & ~ \\
~ & J103846.92+533016.5 & ~ & ~ & ~ & ~ & ~ \\
~ & J103846.64+533011.6 & ~ & ~ & ~ & ~ & ~ \\
~ & J103844.75+533003.0 & ~ & ~ & ~ & ~ & ~ \\
J104522.13+555739.1 & J104522.00+555739.7 & 0.73 $\pm $ 0.01 & 4.92 $\pm $ 0.51 & 3.40 $\pm $ 0.96 & 149.0 & 215.7 \\
~ & J104521.84+555741.6 & ~ & ~ & ~ & ~ & ~ \\
J105114.29+055024.1 & J105114.13+055033.2 & 0.02 $\pm $ 0.002 & 0.51 $\pm $ 0.05 & 0.12 $\pm $ 0.05 & 39.2 & 165.0 \\
J105116.29+055011.8 & J105115.39+055012.5 & 0.005 $\pm $ 0.002 & 0.31 $\pm $ 0.09 & 0.09 $\pm $ 0.05 & 15.2 & 51.1 \\
J105421.65+271425.7 & J105422.19+271426.9 & 0.05 $\pm $ 0.002 & 1.68 $\pm $ 0.22 & 0.31 $\pm $ 0.06 & 30.3 & 164.5 \\
~ & J105421.87+271422.3 & ~ & ~ & ~ & ~ & ~ \\
J112644.26+590920.0 & J112644.21+590921.3 & 0.42 $\pm $ 0.01 & 4.52 $\pm $ 0.52 & 0.96 $\pm $ 0.29 & 93.9 & 442.0 \\
J113835.68+575227.2 & J113835.65+575227.1 & 0.05 $\pm $ 0.002 & 0.74 $\pm $ 0.07 & 0.30 $\pm $ 0.11 & 66.2 & 164.0 \\
J115002.73+150124.0 & J115002.70+150123.2 & 0.06 $\pm $ 0.002 & 1.12 $\pm $ 0.05 & 2.23 $\pm $ 0.31 & 52.1 & 26.1 \\
J115237.27-022809.4 & J115237.20-022810.0 & 0.10 $\pm $ 0.002 & a & 0.12 $\pm $ 0.05 & a & 825.0 \\
J121554.37+130858.0 & J121554.39+130858.4 & 0.04 $\pm $ 0.002 & 1.91 $\pm $ 0.22 & 1.96 $\pm $ 0.41 & 23.2 & 22.6 \\
~ & J121554.48+130900.7 & ~ & ~ & ~ & ~ & ~ \\
~ & J121554.31+130857.4 & ~ & ~ & ~ & ~ & ~ \\
~ & J121554.32+130855.2 & ~ & ~ & ~ & ~ & ~ \\
~ & J121554.68+130856.6 & ~ & ~ & ~ & ~ & ~ \\
J121617.07+693041.6 & J121617.11+693041.6 & 0.10 $\pm $ 0.002 & a & 0.40 $\pm $ 0.19 & a & 247.0 \\
J121635.62+692801.2 & J121635.55+692800.7 & 0.002 b & 3.55 $\pm $ 0.24 & 5.32 $\pm $ 2.08 & 0.5 & 0.4 \\
J121638.17+692849.2 & J121637.89+692847.8 & 0.002 $\pm $ 0.001 & 0.14 $\pm $ 0.03 & 0.55 $\pm $ 0.25 & 16.0 & 4.0 \\
J121640.18+692857.7 & J121639.92+692902.7 & 0.01 $\pm $ 0.001 & 0.67 $\pm $ 0.15 & 0.24 $\pm $ 0.11 & 13.2 & 36.7 \\
J122436.15+392258.7 & J122435.62+392302.9 & 0.46 $\pm $ 0.01 & 4.55 $\pm $ 0.82 & 0.80 $\pm $ 0.32 & 100.7 & 572.5 \\
J122810.74+440649.6 & J122810.95+440648.4 & 0.06 $\pm $ 0.003 & a & 0.86 $\pm $ 0.33 & a & 73.5 \\
J122811.12+440537.1 & J122811.86+440539.1 & 0.68 $\pm $ 0.01 & 1.76 $\pm $ 0.11 & 0.63 $\pm $ 0.21 & 387.8 & 1,083.3 \\
~ & J122811.50+440538.2 & ~ & ~ & ~ & ~ & ~ \\
J122813.81+440710.3 & J122813.85+440710.4 & 0.28 $\pm $ 0.01 & 3.53 $\pm $ 0.24 & 1.74 $\pm $ 0.32 & 80.2 & 162.8 \\
J123028.32+413939.8 & J123028.15+413939.3 & 0.27 $\pm $ 0.01 & a & 0.66 $\pm $ 0.28 & a & 413.9 \\
J123029.48+413927.1 & J123029.56+413926.7 & 0.38 $\pm $ 0.01 & 19.2 $\pm $ 4.8 & 8.20 $\pm $ 0.91 & 19.6 & 45.8 \\
~ & J123029.48+413928.1 & ~ & ~ & ~ & ~ & ~ \\
~ & J123029.31+413928.1 & ~ & ~ & ~ & ~ & ~ \\
~ & J123029.15+413928.2 & ~ & ~ & ~ & ~ & ~ \\
J123034.50+413827.2 & J123034.49+413826.0 & 0.64 $\pm $ 0.01 & 2.88 $\pm $ 0.27 & 2.88 $\pm $ 0.94 & 222.6 & 222.6 \\
~ & J123034.53+413833.2 & ~ & ~ & ~ & ~ & ~ \\
J123034.65+413853.1 & J123034.90+413902.3 & 0.73 $\pm $ 0.01 & 3.55 $\pm $ 0.61 & 1.66 $\pm $ 0.24 & 204.5 & 436.7 \\
~ & J123034.04+413849.3 & ~ & ~ & ~ & ~ & ~ \\
~ & J123034.35+413846.4 & ~ & ~ & ~ & ~ & ~ \\
~ & J123034.18+413850.9 & ~ & ~ & ~ & ~ & ~ \\
~ & J123033.60+413844.0 & ~ & ~ & ~ & ~ & ~ \\
~ & J123035.22+413859.0 & ~ & ~ & ~ & ~ & ~ \\
~ & J123035.07+413857.0 & ~ & ~ & ~ & ~ & ~ \\
~ & J123034.46+413843.5 & ~ & ~ & ~ & ~ & ~ \\
~ & J123034.45+413851.6 & ~ & ~ & ~ & ~ & ~ \\
~ & J123034.13+413854.9 & ~ & ~ & ~ & ~ & ~ \\
J123038.01+413813.0 & J123037.73+413758.5 & 0.08 $\pm $ 0.003 & 0.05 $\pm $ 0.01 & 0.26 $\pm $ 0.11 & 1,612.5 & 303.8 \\
J124356.74+321012.7 & J124356.65+321012.4 & 0.13 $\pm $ 0.004 & 3.3 $\pm $ 0.2 & 1.06 $\pm $ 0.31 & 40.3 & 125.6 \\
J124357.74+321010.7 & J124357.61+321009.2 & 0.07 $\pm $ 0.003 & a & 0.33 $\pm $ 0.10 & a & 222.1 \\
~ & J124358.03+321007.7 & ~ & ~ & ~ & ~ & ~ \\
J124517.09+270731.6 & J124516.98+270731.6 & 0.19 $\pm $ 0.004 & 3.71 $\pm $ 0.05 & 1.43 $\pm $ 0.16 & 51.9 & 134.5 \\
~ & J124517.08+270728.9 & ~ & ~ & ~ & ~ & ~ \\
~ & J124517.17+270728.8 & ~ & ~ & ~ & ~ & ~ \\
~ & J124516.76+270730.9 & ~ & ~ & ~ & ~ & ~ \\
J124813.91-031959.6 & J124813.91-032000.4 & 1.57 $\pm $ 0.03 & 12.2 $\pm $ 1.6 & 3.07 $\pm $ 0.37 & 128.6 & 511.1 \\
~ & J124814.17-032000.2 & ~ & ~ & ~ & ~ & ~ \\
~ & J124813.05-031959.9 & ~ & ~ & ~ & ~ & ~ \\
~ & J124813.61-031957.5 & ~ & ~ & ~ & ~ & ~ \\
J125547.63+041831.5 & J125547.64+041832.9 & 0.15 $\pm $ 0.004 & 5.15 $\pm $ 0.42 & 0.52 $\pm $ 0.11 & 29.1 & 288.3 \\
J125551.23+041747.5 & J125550.51+041757.1 & 0.03 $\pm $ 0.002 & a & 0.14 $\pm $ 0.07 & a & 196.4 \\
J125548.92+041814.8 & J125549.54+041806.7 & 0.14 $\pm $ 0.004 & 2.95 $\pm $ 0.37 & 0.19 $\pm $ 0.07 & 47.4 & 735.3 \\
J125900.36+345043.4 & J125900.41+345043.9 & 0.29 $\pm $ 0.01 & 2.74 $\pm $ 0.07 & 2.67 $\pm $ 0.52 & 105.6 & 108.5 \\
~ & J125900.32+345042.3 & ~ & ~ & ~ & ~ & ~ \\
J133700.61-295155.5 & J133700.34-295150.6 & 13.37 $\pm $ 0.16 & 47 $\pm $ 2 & 46.8 $\pm $ 3.77 & 284.4 & 285.6 \\
~ & J133700.53-295154.8 & ~ & ~ & ~ & ~ & ~ \\
~ & J133700.27-295152.8 & ~ & ~ & ~ & ~ & ~ \\
~ & J133700.38-295159.6 & ~ & ~ & ~ & ~ & ~ \\
~ & J133700.80-295155.5 & ~ & ~ & ~ & ~ & ~ \\
~ & J133701.11-295152.3 & ~ & ~ & ~ & ~ & ~ \\
~ & J133700.89-295203.6 & ~ & ~ & ~ & ~ & ~ \\
~ & J133701.06-295158.2 & ~ & ~ & ~ & ~ & ~ \\
~ & J133700.11-295155.4 & ~ & ~ & ~ & ~ & ~ \\
J133701.37-295129.4 & J133701.35-295127.0 & 1.43 $\pm $ 0.03 & 0.22 $\pm $ 0.04 & 0.25 $\pm $ 0.08 & 6,668.6 & 5,734.8 \\
\enddata
\tablenotetext{a}{Corresponding source could not be individually resolved in the convolved 22 GHz map}
\tablenotetext{b}{W4 band magnitude error is missing for this source in the AllWISE Source Catalog}

\end{deluxetable*}

\clearpage

\begin{figure*}[]
    \centering
    \includegraphics[width=120mm]{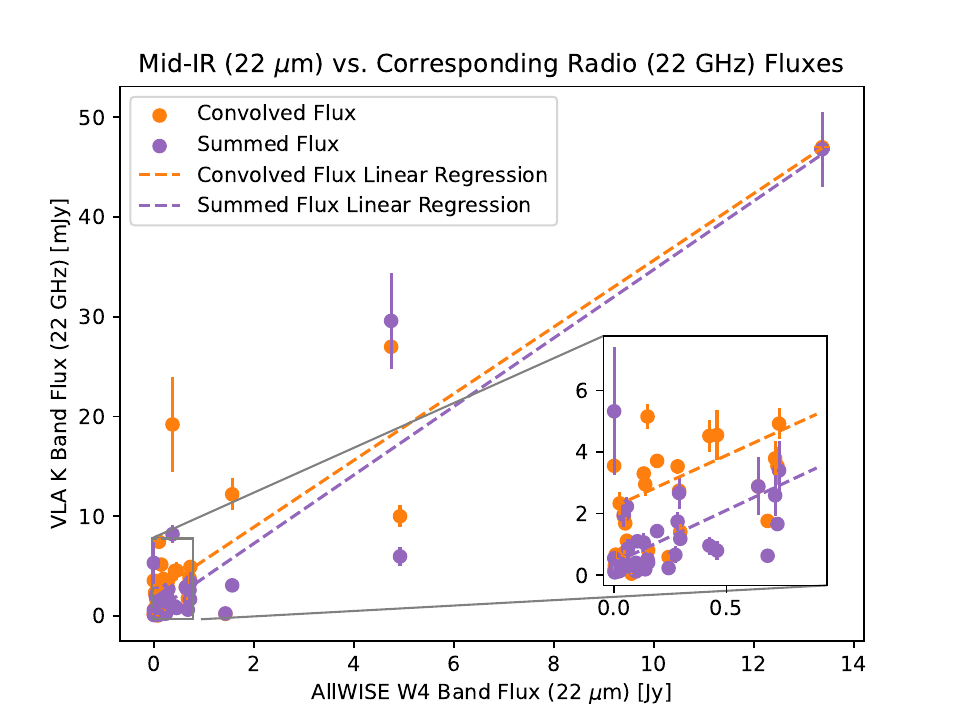}
    \caption{The 22 GHz flux (in mJy) from the convolved VLA K band maps (orange) and from the summed VLA K band maps (purple) vs. the 22 $\mu$m flux (in Jy) from the AllWISE Source Catalog plotted for the 40 corresponding AllWISE Sources. Convolved linear regression: $S_{22 GHz, mJy} \approx 3.34$ $\times S_{22 \mu m, Jy} + 2.22$ Jy, $R^{2}$: 0.802. Summed linear regression: $S_{22 GHz, mJy} \approx 3.43$ $\times S_{22 \mu m, Jy} + 0.39$ Jy, $R^{2}$: 0.854.}
    \label{fig:fig2}
\end{figure*}

\subsection{Future Work} \label{sec:future}
Increasing the sample size of the LWRGS remains the highest priority action moving forward. The LWRGS sample comprises only new VLA observations, allowing for uniformity in the observational and data reduction procedures. Approximately 40 galaxies that were not chosen for the LWRGS sample were temporarily skipped because they already have publicly accessible data in the VLA archive with the appropriate configuration and frequency band combination required for the purposes of this survey. A next paper will apply the same data reduction, interferometric imaging, and flux comparison techniques to these galaxies in the hope of adding more possible SSC candidate regions to our initial list.

It is worth noting that the physical size of these thermal regions spans multiple orders of magnitude, as seen in the column of maximum size estimates in Table 4. Using a combination of new and archival VLA observations of the LWRGS galaxies in higher-resolution configurations (A or B) will allow us to see which thermal regions may contain multiple clusters.

\section{Conclusions}
We present VLA 22GHz observations at 0.95'' resolution of 35 local (d $\lesssim$ 30 Mpc) Wolf-Rayet galaxies. Within this sample:
\begin{enumerate}
    \item We have identified 92 individual regions of high frequency free-free emission. From these sources, we have created a catalog with positions, approximate sizes, peak flux densities, and integrated flux measurements. We have also created a corresponding atlas of VLA 22 GHz maps for these WR galaxy pointings, annotating the 92 individually-identified 22 GHz sources.
    \item We have calculated Lyman continuum rates and derived preliminary estimates of cluster masses and luminosities for the 92 individually-identified 22 GHz sources. All of these sources are found to have Lyman continuum rates at least an order of magnitude greater than those of the most massive Galactic clusters. 39 of the sources meet the $N_{Lyc} > 10^{52} s^{-1}$ criterion expected for an HII region associated with an SSC, and 29 have estimated masses $M \gtrsim 10^{5.5} M_{\odot}$. Thus we conservatively identify at least 29 star-forming regions possibly hosting SSCs, although higher angular resolution observations will be required to confirm the nature of these sources.
    \item We identify two galaxy hosts of likely nonthermal emission, NGC 4236 and UGC 02866. We hypothesize that NGC 4236 a, having an abnormally low mid-IR/radio flux ratio, is a SNR. We also hypothesize that a large percentage of the 22 GHz flux measured for this source is non-thermal synchrotron emission, largely due to its extended 22 GHz emission.
    \item We compare VLA 22 GHz flux measurements with WISE 22 $\mu$m flux measurements. We find wide scatter in the ratios between the two flux measurements, likely a result of the larger beam size ($\sim$12 times larger diameter) associated with the WISE 22 $\mu$m band data. We conclude that using WISE data are not a useful method of detecting individual SSCs within local galaxies but note the potential of the higher-resolution JWST observations to make a more rigorous examination of the mid-IR to radio flux ratio possible.
\end{enumerate}
We present the first stages of the Local Wolf-Rayet Galaxy Survey (LWRGS) and lay the groundwork for the addition of both archival data and new observational data (VLA and ALMA) to this sample.

\begin{acknowledgments}
The National Radio Astronomy Observatory is a facility of the National Science Foundation operated under cooperative agreement by Associated Universities, Inc. This publication makes use of data products from the Wide-field Infrared Survey Explorer, which is a joint project of the University of California, Los Angeles, and the Jet Propulsion Laboratory/California Institute of Technology, funded by the National Aeronautics and Space Administration. This research has made use of the NASA/IPAC Extragalactic Database (NED), which is funded by the National Aeronautics and Space Administration and operated by the California Institute of Technology.
Support for this work
was provided by the National Science Foundation (NSF) through an SOS grant from the National Radio Astronomy Observatory to NF and NSF grant AST 2006433 to JLT.
\end{acknowledgments}

\facilities{EVLA,WISE}
\software{CASA \citep{2007ASPC..376..127M}, CARTA \citep{angus_comrie_2018_3377984}, Starburst99 \citep{1999ApJS..123....3L}, SAOImageDS9 \citep{2000ascl.soft03002S}}

\bibliography{cpd}{}

\begin{thebibliography}{}
\expandafter\ifx\csname natexlab\endcsname\relax\def\natexlab#1{#1}\fi
\providecommand{\url}[1]{\href{#1}{#1}}
\providecommand{\dodoi}[1]{doi:~\href{http://doi.org/#1}{\nolinkurl{#1}}}
\providecommand{\doeprint}[1]{\href{http://ascl.net/#1}{\nolinkurl{http://ascl.net/#1}}}
\providecommand{\doarXiv}[1]{\href{https://arxiv.org/abs/#1}{\nolinkurl{https://arxiv.org/abs/#1}}}

\bibitem[{{Alves} \& {Homeier}(2003)}]{2003ApJ...589L..45A}
{Alves}, J., \& {Homeier}, N. 2003, \apjl, 589, L45, \dodoi{10.1086/375801}

\bibitem[{{Beck}(2008)}]{2008A&A...489..567B}
{Beck}, S.~C. 2008, \aap, 489, 567, \dodoi{10.1051/0004-6361:200810441}

\bibitem[{{Brinchmann} {et~al.}(2008){Brinchmann}, {Kunth}, \& {Durret}}]{2008A&A...485..657B}
{Brinchmann}, J., {Kunth}, D., \& {Durret}, F. 2008, \aap, 485, 657, \dodoi{10.1051/0004-6361:200809783}

\bibitem[{{Calzetti} {et~al.}(2015){Calzetti}, {Lee}, {Sabbi}, {Adamo}, {Smith}, {Andrews}, {Ubeda}, {Bright}, {Thilker}, {Aloisi}, {Brown}, {Chandar}, {Christian}, {Cignoni}, {Clayton}, {da Silva}, {de Mink}, {Dobbs}, {Elmegreen}, {Elmegreen}, {Evans}, {Fumagalli}, {Gallagher}, {Gouliermis}, {Grebel}, {Herrero}, {Hunter}, {Johnson}, {Kennicutt}, {Kim}, {Krumholz}, {Lennon}, {Levay}, {Martin}, {Nair}, {Nota}, {{\"O}stlin}, {Pellerin}, {Prieto}, {Regan}, {Ryon}, {Schaerer}, {Schiminovich}, {Tosi}, {Van Dyk}, {Walterbos}, {Whitmore}, \& {Wofford}}]{2015AJ....149...51C}
{Calzetti}, D., {Lee}, J.~C., {Sabbi}, E., {et~al.} 2015, \aj, 149, 51, \dodoi{10.1088/0004-6256/149/2/51}

\bibitem[{Comrie {et~al.}(2018)Comrie, Wang, Hwang, Moraghan, Harris, Pińska, Raul-Omar, Chiang, Chang, Hsu, Pang, Simmonds, Lin, \& Jan}]{angus_comrie_2018_3377984}
Comrie, A., Wang, K.-S., Hwang, Y.-H., {et~al.} 2018, {CARTA: The Cube Analysis and Rendering Tool for Astronomy},  Zenodo, \dodoi{10.5281/zenodo.3377984}

\bibitem[{{Consiglio} {et~al.}(2016){Consiglio}, {Turner}, {Beck}, \& {Meier}}]{2016ApJ...833L...6C}
{Consiglio}, S.~M., {Turner}, J.~L., {Beck}, S., \& {Meier}, D.~S. 2016, \apjl, 833, L6, \dodoi{10.3847/2041-8205/833/1/L6}

\bibitem[{{Conti}(1991)}]{1991ApJ...377..115C}
{Conti}, P.~S. 1991, \apj, 377, 115, \dodoi{10.1086/170340}

\bibitem[{{Conway} {et~al.}(1990){Conway}, {Cornwell}, \& {Wilkinson}}]{1990MNRAS.246..490C}
{Conway}, J.~E., {Cornwell}, T.~J., \& {Wilkinson}, P.~N. 1990, \mnras, 246, 490

\bibitem[{Crowther(2007)}]{doi:10.1146/annurev.astro.45.051806.110615}
Crowther, P.~A. 2007, Annual Review of Astronomy and Astrophysics, 45, 177, \dodoi{10.1146/annurev.astro.45.051806.110615}

\bibitem[{{Cutri} {et~al.}(2012){Cutri}, {Wright}, {Conrow}, {Bauer}, {Benford}, {Brandenburg}, {Dailey}, {Eisenhardt}, {Evans}, {Fajardo-Acosta}, {Fowler}, {Gelino}, {Grillmair}, {Harbut}, {Hoffman}, {Jarrett}, {Kirkpatrick}, {Leisawitz}, {Liu}, {Mainzer}, {Marsh}, {Masci}, {McCallon}, {Padgett}, {Ressler}, {Royer}, {Skrutskie}, {Stanford}, {Wyatt}, {Tholen}, {Tsai}, {Wachter}, {Wheelock}, {Yan}, {Alles}, {Beck}, {Grav}, {Masiero}, {McCollum}, {McGehee}, {Papin}, \& {Wittman}}]{2012wise.rept....1C}
{Cutri}, R.~M., {Wright}, E.~L., {Conrow}, T., {et~al.} 2012, {Explanatory Supplement to the WISE All-Sky Data Release Products}, Explanatory Supplement to the WISE All-Sky Data Release Products

\bibitem[{{De Pree} {et~al.}(2000){De Pree}, {Wilner}, {Goss}, {Welch}, \& {McGrath}}]{2000ApJ...540..308D}
{De Pree}, C.~G., {Wilner}, D.~J., {Goss}, W.~M., {Welch}, W.~J., \& {McGrath}, E. 2000, \apj, 540, 308, \dodoi{10.1086/309315}

\bibitem[{{de Vaucouleurs} {et~al.}(1991){de Vaucouleurs}, {de Vaucouleurs}, {Corwin}, {Buta}, {Paturel}, \& {Fouque}}]{1991rc3..book.....D}
{de Vaucouleurs}, G., {de Vaucouleurs}, A., {Corwin}, Herold~G., J., {et~al.} 1991, {Third Reference Catalogue of Bright Galaxies}

\bibitem[{{Gorjian} {et~al.}(2001){Gorjian}, {Turner}, \& {Beck}}]{2001ApJ...554L..29G}
{Gorjian}, V., {Turner}, J.~L., \& {Beck}, S.~C. 2001, \apjl, 554, L29, \dodoi{10.1086/320923}

\bibitem[{{Hannon} {et~al.}(2022){Hannon}, {Lee}, {Whitmore}, {Mobasher}, {Thilker}, {Chandar}, {Adamo}, {Wofford}, {Orozco-Duarte}, {Calzetti}, {Della Bruna}, {Kreckel}, {Groves}, {Barnes}, {Boquien}, {Belfiore}, \& {Linden}}]{2022MNRAS.512.1294H}
{Hannon}, S., {Lee}, J.~C., {Whitmore}, B.~C., {et~al.} 2022, \mnras, 512, 1294, \dodoi{10.1093/mnras/stac550}

\bibitem[{{Hurley-Walker} {et~al.}(2009){Hurley-Walker}, {Scaife}, {Green}, {Davies}, {Grainge}, {Hobson}, {Jones}, {Kaneko}, {Lasenby}, {Pooley}, {Saunders}, {Scott}, {Titterington}, {Waldram}, \& {Zwart}}]{2009MNRAS.396..365H}
{Hurley-Walker}, N., {Scaife}, A.~M.~M., {Green}, D.~A., {et~al.} 2009, \mnras, 396, 365, \dodoi{10.1111/j.1365-2966.2009.14583.x}

\bibitem[{{Leitherer} {et~al.}(1999){Leitherer}, {Schaerer}, {Goldader}, {Delgado}, {Robert}, {Kune}, {de Mello}, {Devost}, \& {Heckman}}]{1999ApJS..123....3L}
{Leitherer}, C., {Schaerer}, D., {Goldader}, J.~D., {et~al.} 1999, \apjs, 123, 3, \dodoi{10.1086/313233}

\bibitem[{{Linden} {et~al.}(2020){Linden}, {Murphy}, {Dong}, {Momjian}, {Kennicutt}, {Meier}, {Schinnerer}, \& {Turner}}]{2020ApJS..248...25L}
{Linden}, S.~T., {Murphy}, E.~J., {Dong}, D., {et~al.} 2020, \apjs, 248, 25, \dodoi{10.3847/1538-4365/ab8a4d}

\bibitem[{{Mahajan} {et~al.}(2019){Mahajan}, {Ashby}, {Willner}, {Barmby}, {Fazio}, {Maragkoudakis}, {Raychaudhury}, \& {Zezas}}]{2019MNRAS.482..560M}
{Mahajan}, S., {Ashby}, M.~L.~N., {Willner}, S.~P., {et~al.} 2019, \mnras, 482, 560, \dodoi{10.1093/mnras/sty2699}

\bibitem[{{McMullin} {et~al.}(2007){McMullin}, {Waters}, {Schiebel}, {Young}, \& {Golap}}]{2007ASPC..376..127M}
{McMullin}, J.~P., {Waters}, B., {Schiebel}, D., {Young}, W., \& {Golap}, K. 2007, in Astronomical Society of the Pacific Conference Series, Vol. 376, Astronomical Data Analysis Software and Systems XVI, ed. R.~A. {Shaw}, F.~{Hill}, \& D.~J. {Bell}, 127

\bibitem[{{Melnick}(1979)}]{1979ApJ...228..112M}
{Melnick}, J. 1979, \apj, 228, 112, \dodoi{10.1086/156827}

\bibitem[{{Moln{\'a}r} {et~al.}(2021){Moln{\'a}r}, {Sargent}, {Leslie}, {Magnelli}, {Schinnerer}, {Zamorani}, {Delhaize}, {Smol{\v{c}}i{\'c}}, {Tisani{\'c}}, \& {Vardoulaki}}]{2021MNRAS.504..118M}
{Moln{\'a}r}, D.~C., {Sargent}, M.~T., {Leslie}, S., {et~al.} 2021, \mnras, 504, 118, \dodoi{10.1093/mnras/stab746}

\bibitem[{{Murphy} {et~al.}(2018){Murphy}, {Dong}, {Momjian}, {Linden}, {Kennicutt}, {Meier}, {Schinnerer}, \& {Turner}}]{2018ApJS..234...24M}
{Murphy}, E.~J., {Dong}, D., {Momjian}, E., {et~al.} 2018, \apjs, 234, 24, \dodoi{10.3847/1538-4365/aa99d7}

\bibitem[{{Natta} \& {Panagia}(1984)}]{1984ApJ...287..228N}
{Natta}, A., \& {Panagia}, N. 1984, \apj, 287, 228, \dodoi{10.1086/162681}

\bibitem[{{Oni{\'c}} \& {Uro{\v{s}}evi{\'c}}(2015)}]{2015ApJ...805..119O}
{Oni{\'c}}, D., \& {Uro{\v{s}}evi{\'c}}, D. 2015, \apj, 805, 119, \dodoi{10.1088/0004-637X/805/2/119}

\bibitem[{{Rau} \& {Cornwell}(2011)}]{2011A&A...532A..71R}
{Rau}, U., \& {Cornwell}, T.~J. 2011, \aap, 532, A71, \dodoi{10.1051/0004-6361/201117104}

\bibitem[{{Sault} \& {Wieringa}(1994)}]{1994A&AS..108..585S}
{Sault}, R.~J., \& {Wieringa}, M.~H. 1994, \aaps, 108, 585

\bibitem[{{Schaerer} {et~al.}(1999){Schaerer}, {Contini}, \& {Pindao}}]{1999A&AS..136...35S}
{Schaerer}, D., {Contini}, T., \& {Pindao}, M. 1999, \aaps, 136, 35, \dodoi{10.1051/aas:1999197}

\bibitem[{Silich {et~al.}(2020)Silich, Tenorio-Tagle, Martínez-González, \& Turner}]{10.1093/mnras/staa705}
Silich, S., Tenorio-Tagle, G., Martínez-González, S., \& Turner, J. 2020, Monthly Notices of the Royal Astronomical Society, 494, 97, \dodoi{10.1093/mnras/staa705}

\bibitem[{{Silich} {et~al.}(2004){Silich}, {Tenorio-Tagle}, \& {Rodr{\'\i}guez-Gonz{\'a}lez}}]{2004ApJ...610..226S}
{Silich}, S., {Tenorio-Tagle}, G., \& {Rodr{\'\i}guez-Gonz{\'a}lez}, A. 2004, \apj, 610, 226, \dodoi{10.1086/421702}

\bibitem[{{Smithsonian Astrophysical Observatory}(2000)}]{2000ascl.soft03002S}
{Smithsonian Astrophysical Observatory}. 2000, {SAOImage DS9: A utility for displaying astronomical images in the X11 window environment}, Astrophysics Source Code Library, record ascl:0003.002

\bibitem[{{Teh} {et~al.}(2023){Teh}, {Grasha}, {Krumholz}, {Battisti}, {Calzetti}, {Rousseau-Nepton}, {Rhea}, {Adamo}, {Kennicutt}, {Grebel}, {Cook}, {Combes}, {Messa}, {Linden}, {Klessen}, {Vilchez}, {Fumagalli}, {McLeod}, {Smith}, {Chemin}, {Wang}, {Sabbi}, {Sacchi}, {Petric}, {Della Bruna}, \& {Boselli}}]{2023MNRAS.524.1191T}
{Teh}, J.~W., {Grasha}, K., {Krumholz}, M.~R., {et~al.} 2023, \mnras, 524, 1191, \dodoi{10.1093/mnras/stad1780}

\bibitem[{{Tenorio-Tagle} {et~al.}(2015){Tenorio-Tagle}, {Mu{\~n}oz-Tu{\~n}{\'o}n}, {Silich}, \& {Cassisi}}]{2015ApJ...814L...8T}
{Tenorio-Tagle}, G., {Mu{\~n}oz-Tu{\~n}{\'o}n}, C., {Silich}, S., \& {Cassisi}, S. 2015, \apjl, 814, L8, \dodoi{10.1088/2041-8205/814/1/L8}

\bibitem[{{Turner} {et~al.}(2017){Turner}, {Consiglio}, {Beck}, {Goss}, {Ho}, {Meier}, {Silich}, \& {Zhao}}]{2017ApJ...846...73T}
{Turner}, J.~L., {Consiglio}, S.~M., {Beck}, S.~C., {et~al.} 2017, \apj, 846, 73, \dodoi{10.3847/1538-4357/aa8669}

\bibitem[{{Turner} \& {Ho}(1994)}]{1994ApJ...421..122T}
{Turner}, J.~L., \& {Ho}, P. T.~P. 1994, \apj, 421, 122, \dodoi{10.1086/173631}

\bibitem[{{Vollmer} {et~al.}(2022){Vollmer}, {Soida}, \& {Dallant}}]{2022A&A...667A..30V}
{Vollmer}, B., {Soida}, M., \& {Dallant}, J. 2022, \aap, 667, A30, \dodoi{10.1051/0004-6361/202142877}

\bibitem[{{Wright} {et~al.}(2010){Wright}, {Eisenhardt}, {Mainzer}, {Ressler}, {Cutri}, {Jarrett}, {Kirkpatrick}, {Padgett}, {McMillan}, {Skrutskie}, {Stanford}, {Cohen}, {Walker}, {Mather}, {Leisawitz}, {Gautier}, {McLean}, {Benford}, {Lonsdale}, {Blain}, {Mendez}, {Irace}, {Duval}, {Liu}, {Royer}, {Heinrichsen}, {Howard}, {Shannon}, {Kendall}, {Walsh}, {Larsen}, {Cardon}, {Schick}, {Schwalm}, {Abid}, {Fabinsky}, {Naes}, \& {Tsai}}]{2010AJ....140.1868W}
{Wright}, E.~L., {Eisenhardt}, P. R.~M., {Mainzer}, A.~K., {et~al.} 2010, \aj, 140, 1868, \dodoi{10.1088/0004-6256/140/6/1868}

\end{thebibliography}
\bibliographystyle{aasjournal}

\begin{appendix}
\begin{figure*}[!ht]
    \centering
    \includegraphics[width=100mm,scale=0.5]{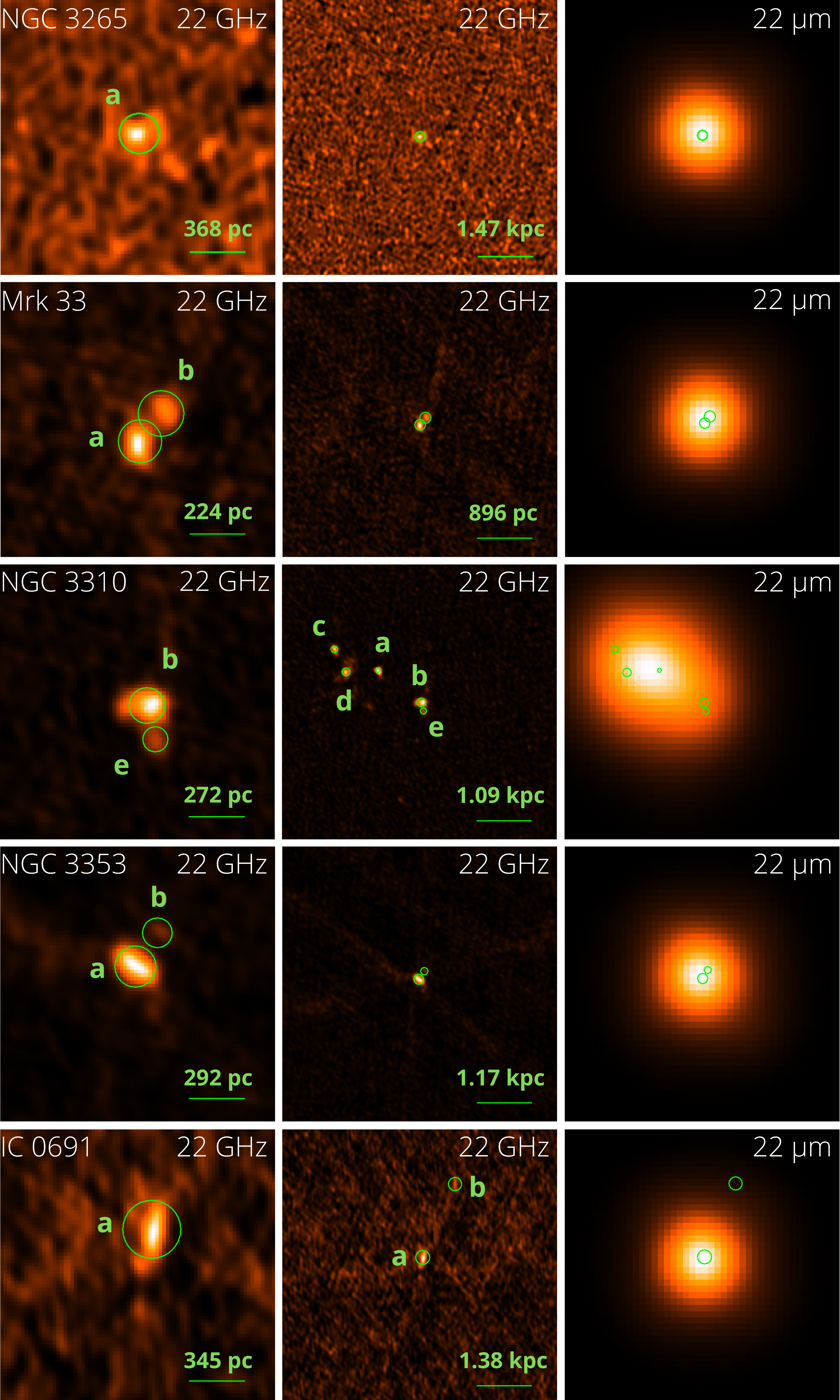}
    \caption{See Figure 1 for description.}
    \label{fig:a2}
\end{figure*}

\begin{figure*}[!ht]
    \centering
    \includegraphics[width=100mm,scale=0.5]{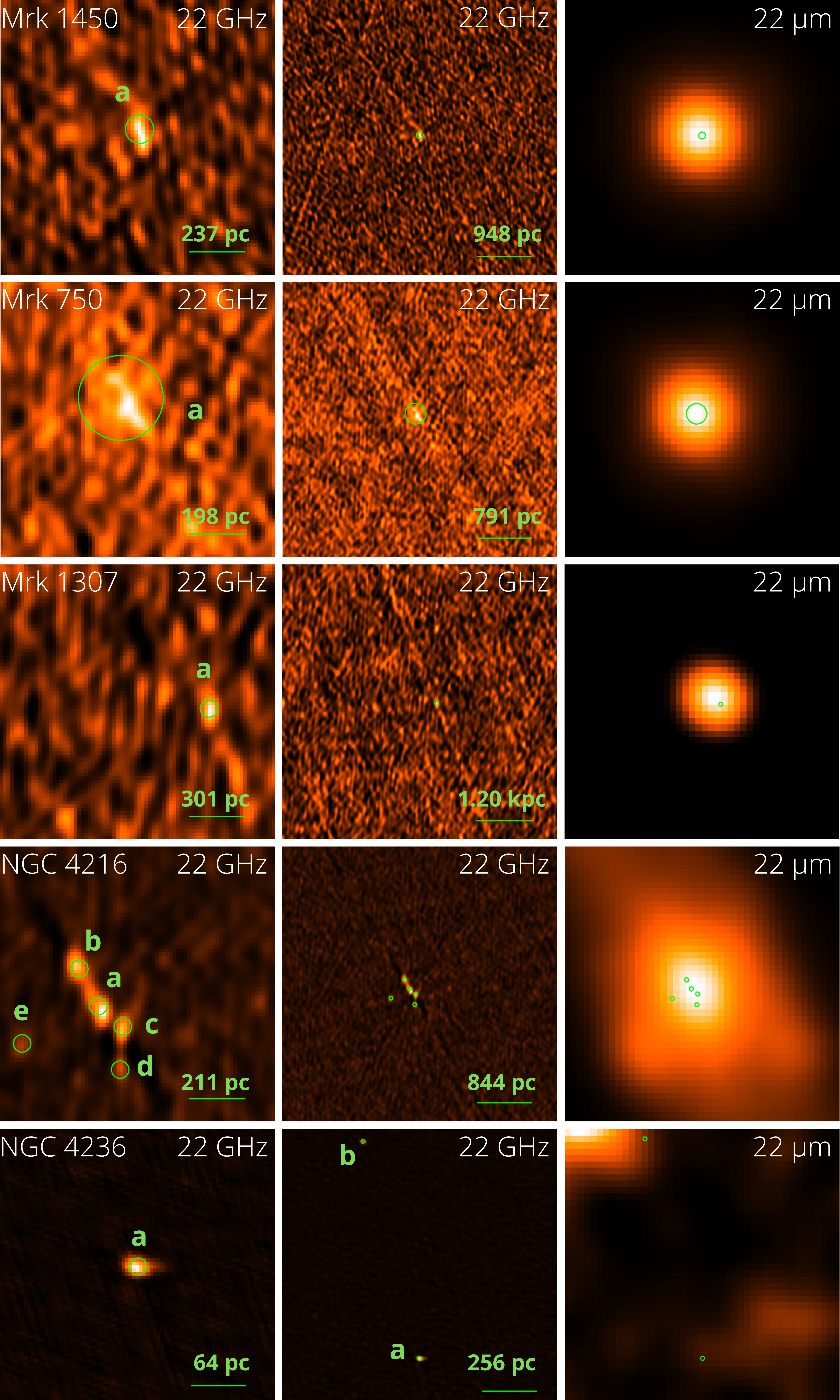}
    \caption{See Figure 1 for description.}
    \label{fig:a3}
\end{figure*}

\begin{figure*}[!ht]
    \centering
    \includegraphics[width=100mm,scale=0.5]{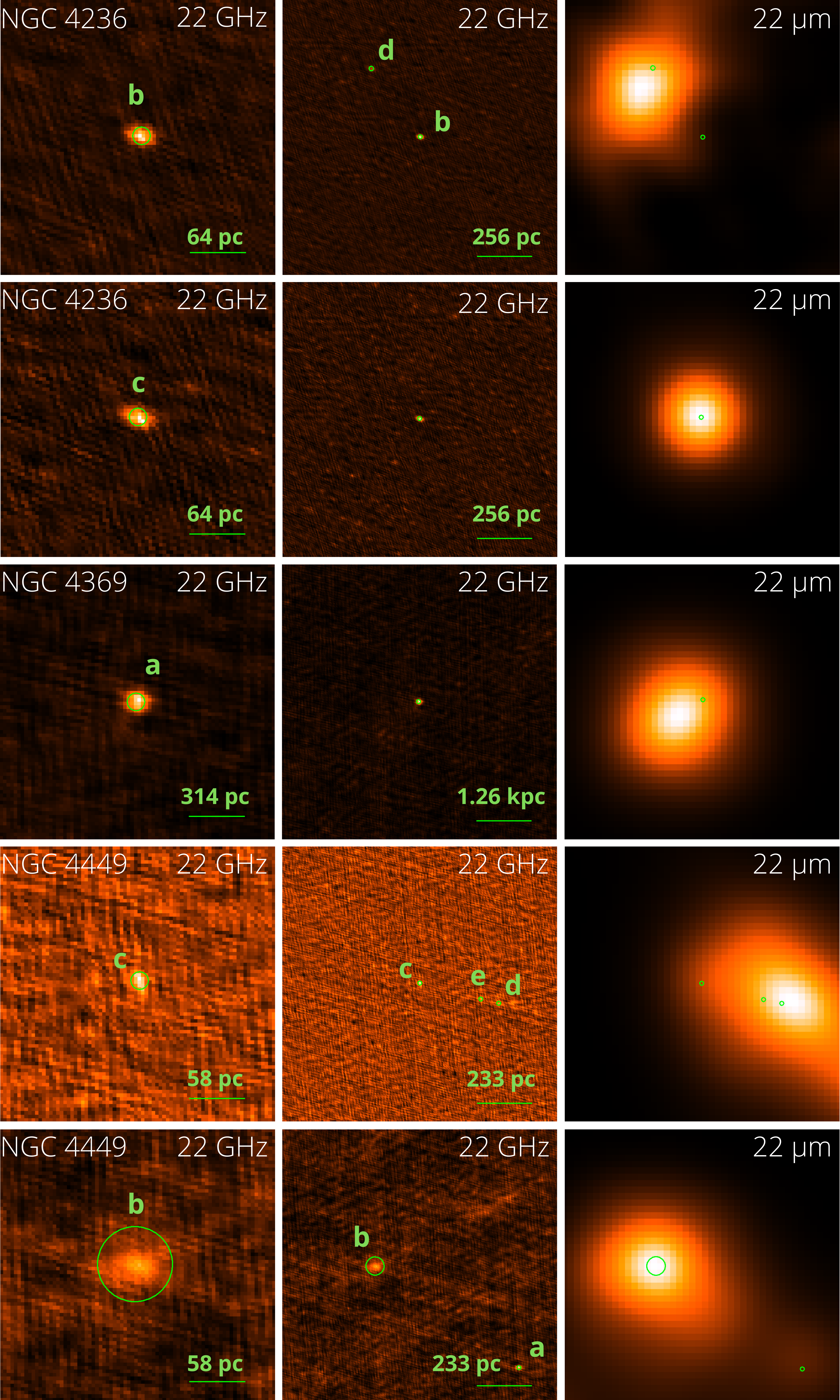}
    \caption{See Figure 1 for description.}
    \label{fig:a4}
\end{figure*}

\begin{figure*}[!ht]
    \centering
    \includegraphics[width=100mm,scale=0.5]{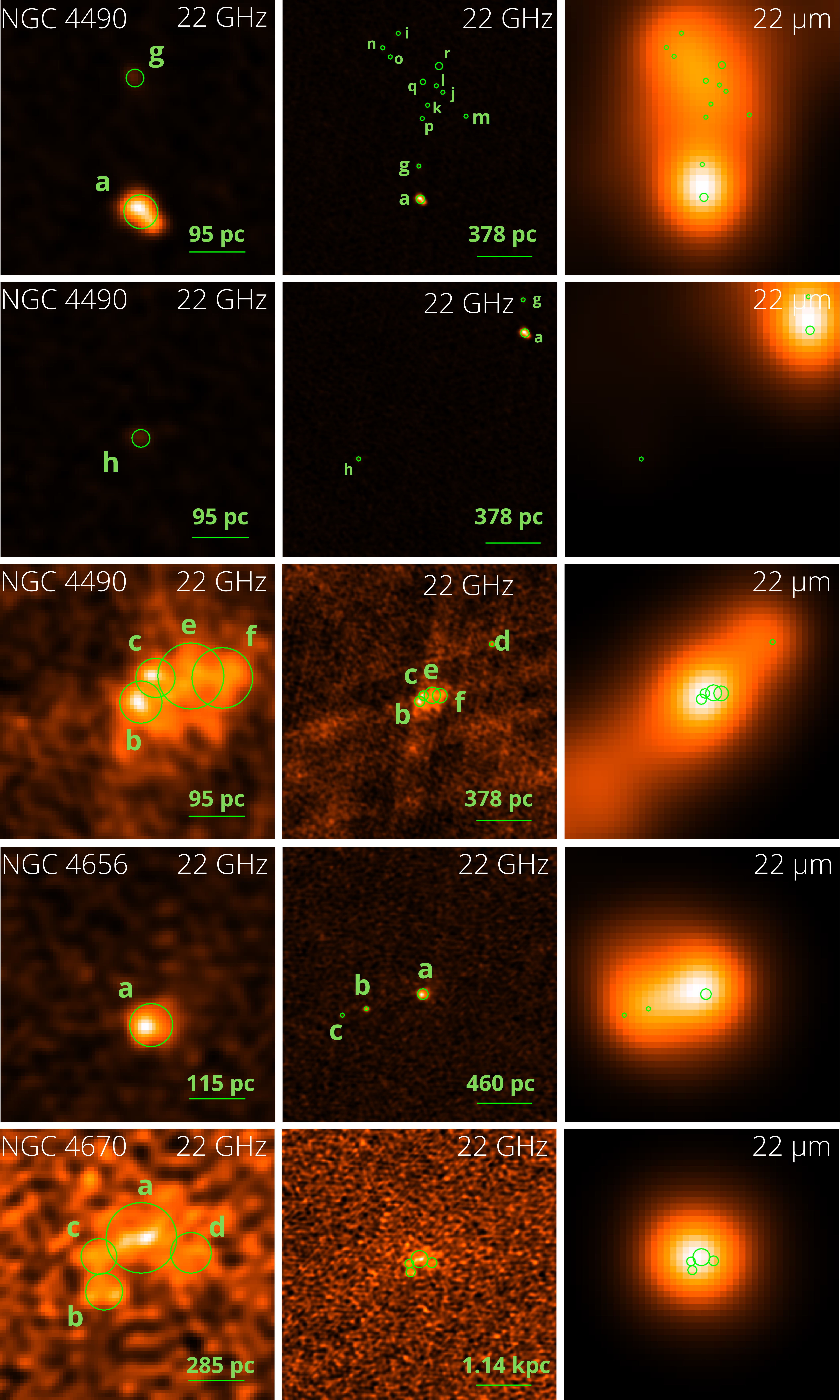}
    \caption{See Figure 1 for description.}
    \label{fig:a5}
\end{figure*}

\begin{figure*}[!ht]
    \centering
    \includegraphics[width=100mm,scale=0.5]{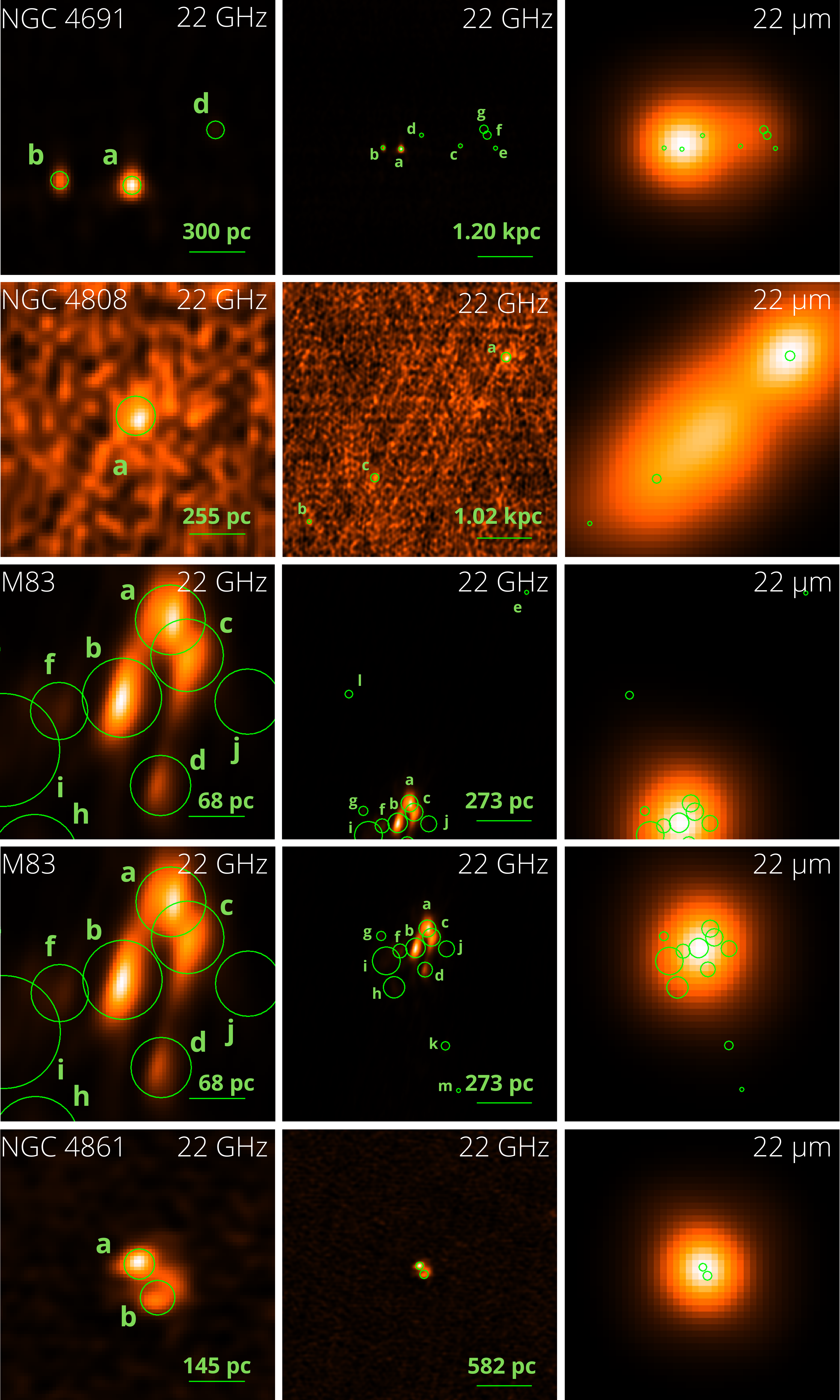}
    \caption{See Figure 1 for description.}
    \label{fig:a6}
\end{figure*}

\begin{figure*}[!ht]
    \centering
    \includegraphics[width=100mm,scale=0.5]{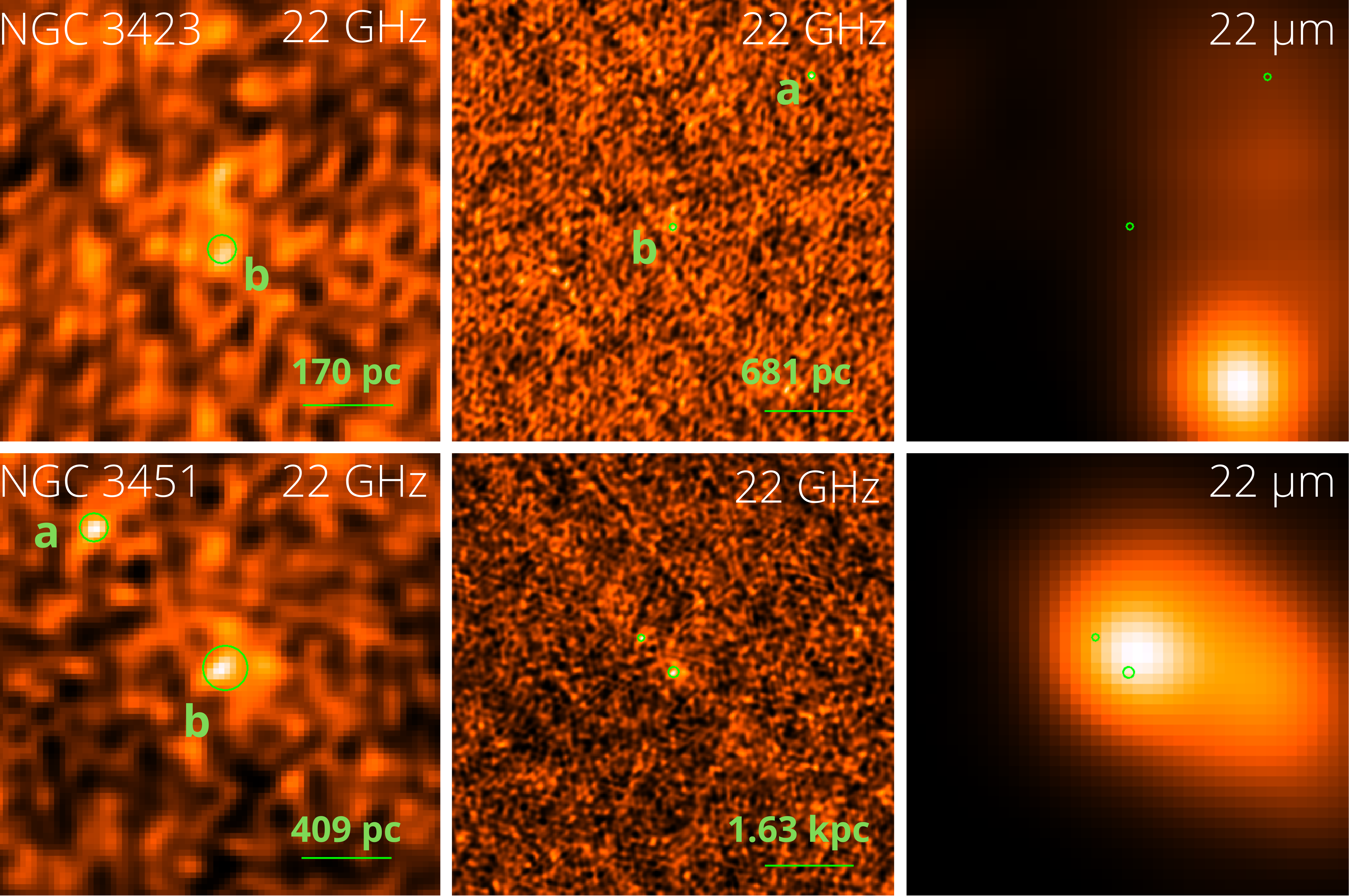}
    \caption{See Figure 1 for description.}
    \label{fig:a7}
\end{figure*}

\end{appendix}
\end{document}